\documentclass[a4paper,fleqn,usenatbib]{mnras}
\usepackage{amssymb}
\usepackage{amsmath}
\usepackage{graphicx}
\usepackage{natbib}
\usepackage{hyperref}
\usepackage{pdflscape}
\usepackage{longtable}
\usepackage{multicol}
\usepackage{multirow}
\usepackage{url}
\usepackage{graphicx}
\usepackage{color}
\usepackage{txfonts}

\DeclareRobustCommand{\ION}[2]{%
\relax\ifmmode
\ifx\testbx\f@series
{\mathbf{#1\,\mathsc{#2}}}\else
{\mathrm{#1\,\mathsc{#2}}}\fi
\else\textup{#1\,{\mdseries\textsc{#2}}}%
\fi}

\newcommand{\Hii}{\ION{H}{ii}}
\newcommand{\hii}{\ION{H}{ii}}

\newcommand{\nii}{[\ION{N}{ii}]}
\newcommand{\oi}{[\ION{O}{i}]}

\newcommand{\oiii}{[\ION{O}{iii}]}
\newcommand{\sii}{[\ION{S}{ii}]}

\newcommand{\Ha}{$\rm{H}\alpha$}
\newcommand{\Hb}{$\rm{H}\beta$}

\newcommand{\pyHII}{{\sc pyHIIextractor}}
\newcommand{\pipe}{{\sc Pipe3D}}
\newcommand{\orcid}[1]{\href{https://orcid.org/#1}{\includegraphics[width=16pt]{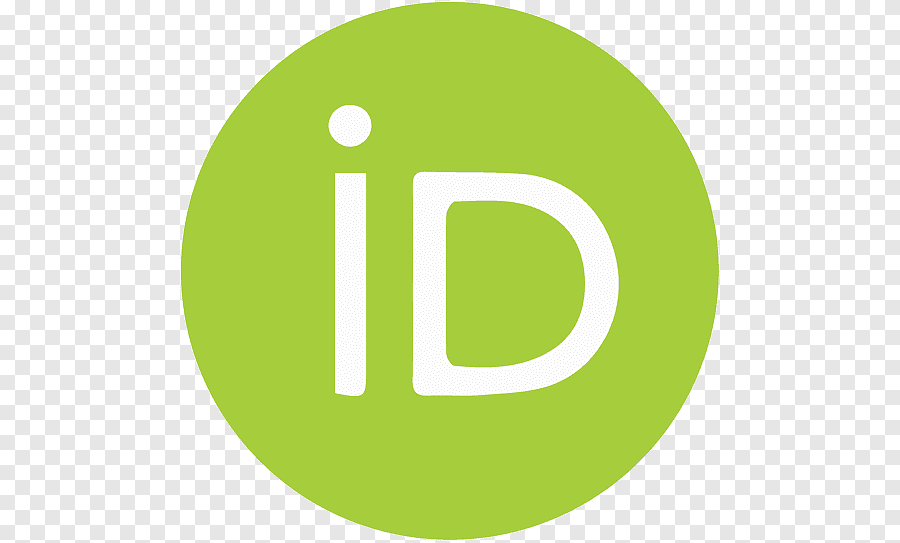}}}

\title[\hii\ regions and diffuse ionized gas maps in the AMUSING++ compilation]{\hii\ regions and diffuse ionized gas in the AMUSING++ Compilation: \\
I. Catalogue presentation}

\author[A.Z.Lugo-Aranda]{
A.Z. Lugo-Aranda$^{1}$,
S.F. S\'anchez$^{1}$,
J.K. Barrera-Ballesteros$^{1}$,
C. L\'opez-Cob\'a$^{2}$\orcid{https://orcid.org/0000-0003-1045-0702},
C. Espinosa-Ponce$^{1}$\newauthor
L. Galbany$^{3,4}$\orcid{https://orcid.org/0000-0002-1296-6887},
Joseph P. Anderson$^{5,6}$\orcid{https://orcid.org/0000-0003-0227-3451}
\\
$^1$Instituto de Astronom\'ia, Universidad Nacional Aut\'onoma de M\'exico, A.~P. 70-264, C.P. 04510, M\'exico, D.F., Mexico.\\
$^{2}$Institute of Astronomy and Astrophysics, Academia Sinica, No. 1, Section 4, Roosevelt Road, Taipei 10617, Taiwan.\\
$^{3}$Institute of Space Sciences (ICE-CSIC), Campus UAB, Carrer de Can Magrans, s/n, E-08193 Barcelona, Spain.\\
$^{4}$Institut d’Estudis Espacials de Catalunya (IEEC), E-08034 Barcelona, Spain.\\
$^{5}$European Southern Observatory, Alonso de C\'ordova 3107, Casilla 19, Santiago, Chile.\\
$^{6}$Millennium Institute of Astrophysics MAS, Nuncio Monsenor Sotero Sanz 100, Off. 104, Providencia, Santiago, Chile.\\
}

\date{Accepted XXX. Received YYY; in original form ZZZ}

\pubyear{2024}

\begin{document}
\label{firstpage}
\pagerange{\pageref{firstpage}--\pageref{lastpage}}
\maketitle

\begin{abstract}

We present a catalog of $\sim$52,000 extragalactic \hii\ regions and their spectroscopic properties obtained using Integral Field Spectroscopy (IFS) from MUSE observations. The sample analyzed in this study contains 678 galaxies within the nearby Universe (0.004 $< z <$ 0.06) covering different morphological types and a wide range of stellar masses (6 $<$ log(M$_{*}$/M$_{\odot}$) $<$ 13). Each galaxy was analyzed using the \pipe\ and \pyHII\ codes to obtain information of the ionized gas and underlying stellar populations. Specifically, the fluxes, equivalent widths, velocities and velocity dispersions of 30 emission lines covering the wavelength range between $\lambda$4750$\AA$ to $\lambda$9300$\AA$, were extracted and were used to estimate luminosity weighted ages and metallicities of the underlying stellar populations from each \hii\ region (of the original sample we detect \hii\ regions in 539 galaxies). In addition, we introduce and apply a novel method and independent of any intrinsic physical property to estimate and decontaminate the contribution of the diffuse ionized gas. Using the final catalog, we explore the dependence of properties of the \hii\ regions on different local and global galaxy parameters: (i) Hubble type, (ii) stellar mass, (iii) galactocentric distance, and (iv) the age and metallicity of the underlying/neighbour stellar populations.
We confirm known relations between properties of the \hii\ regions and the underlying stellar populations (in particular with the age) uncovered using data of lower spatial and spectral resolution. 
Furthermore, we describe the existence of two main families of diffuse ionized gas different for galaxies host or not of \hii\ regions.

\end{abstract}


\section{Introduction}
\label{sec:intro}

\hii\ regions are gas clouds, gravitationally or pressure bonded, ionized by young massive OB stars. The short-life of the low-mass populations of young clusters and associations \citep[$<$10 Myr,][]{wrightgoodwinjeffries2022} defines the time-scale of these regions, being the reason why \hii\ regions are essential tracers of recent star-formation processes in galaxies \citep[e.g.,][]{relanokennicutt2009, sanchezperezrosalesortega2015}. Being very bright objects, that present strong ionized gas emission lines, their properties have been explored for decades 
(e.g., \citealp{peimbertbatiz1960}, \citealp{peimbertcostero1961}, \citealp{odellpeimbertkinman1964}). They present a wide range of sizes, (from a few pc e.g., 8 pc for Orion nebula, \citealp{ander14}; to kpc-scales, e.g., $\sim$1 kpc for NGC 5471, \citealp{garciaperezdiaz2011}), electron densities \citep[10-1000 cm$^{-3}$][]{hunthirashita2009}, and electron temperatures \citep[4000$^{\circ}$K and 10000$^{\circ}$K][]{shaver1970}. Being mostly composed by Hydrogen and Helium, their metal content is the consequence of the chemical enrichment history at the location in which they are formed, sharing their metal composition with the stars that ionize them, affected by the influence of additional global factors, e.g., outflows \citep{qinwangzhao2008}, inflows \citep{watkinskreckelgroves2023}, mergers \citep{ferreiropastorizarickes2008} and bars \citep{zuritaperez2008}. They are one of the fundamental tools to explore the current chemical content of the interstellar medium in galaxies \citep[e.g.][]{searle1971,rollestonsmartt2000,carigipeimbertpeimbert2019}, and their evolution along cosmological times. 

\hii\ regions have been used for decades to explore the abundance gradients in galaxies \citep[e.g.,][]{ searle1971, vilacostas1992, bresolin2019}, being fundamental tracers of (i) the inside-out growth of intermediate/massive spiral galaxies ($>$10$^{9}$M$_\odot$) \citep{matteucci1989, boissierprantzos2000}, (ii) the effects induced by bars in galaxy evolution \citep{roskardebattistastinson2008,zuritafloridobresolin2021}, (iii) gas migration in galaxies \citep{sanchezgalbanyperez2015census}, and (iv) the details of the SF processes itself \citep{kennicuttedgarhodge1989,bradleyknapenbeckman2006}. The advent of new observational techniques, in particular wide-field integral-field spectroscopy applied to large number of galaxies \citep{rosalesortegakennicuttsanchez2010,blancweinirlsong2013} has allowed new explorations of the properties of \hii\ regions \citep{marinogildepazcastillo2012}. More recently, the introduction of intermediate spatial-resolution but large IFS surveys, such as CALIFA (\citealp{califapresentation}), and high spatial-resolution but more statistically limited ones, such PHANGS-MUSE \citep{emsellemschinnerersantoro2022}, have allowed to obtain large and detailed catalogs of the spectroscopic properties of these objects with un-precedent detail \citep{espi20,groveskreckelsantoro2023}. These catalogs has uncovered fundamental relations between the local properties such as mass and metallicity \citep{rosalesortegasancheziglesias2012}, patterns connecting them with those of their underlying stellar populations \citep{sanchezperezrosalesortega2015,espi20}, allowing detailed explorations on the nature of the diffuse-ionized gas \citep{belfioresantorogroves2022}.

In this work we present the largest available catalog of spectroscopic properties of \hii\ regions ($\sim$52,000 individual regions) using IFS MUSE observations \citep{muse}, extracted from the AMUSING++ compilation of nearby Universe galaxies ($z<$0.1, \citealp{lopezcobasanchez2020}). The catalog includes the parameters recovered by the Pipe3D pipeline \citep{sanchezperezsanchez2016fit3d,sanchezperezsanchez2016,lacerdasanchezmejia2022}, from these dataset ($\sim$700 galaxies), comprising both ionized gas and underlying stellar population properties. The high spatial physical-resolution ($\sim$0.1-1 kpc) of these data and the introduction of a novel technique to select and segregate the \hii\ regions (\pyHII, \citealp{lugosanchezespinosa2022}), allow us to perform a detailed separation and decontamination of the diffuse-ionized gas (DIG) component. We use this catalog to continue previous explorations performed using lower spatial-resolution IFS data \citep[e.g., CALIFA][]{espi20, espinosaponcesanchezmorisset2022}, confirming (i) the reported patterns across the classical diagnostic diagrams and (ii) the relation between the properties of \hii\ regions and those of the underlying stellar populations. Furthermore, the new dataset has allowed us to explore the different natures of the DIG in galaxies hosting and not hosting \hii\ regions.

The structure of this article is as follows: Section \ref{sec:sample} presents the main characteristics of the MUSE instrument, describing the properties of the selected galaxy compilation, and presenting a qualitative comparison with the properties of a diameter selected sample including redshift, effective radius and star formation rate (SFR); Section \ref{sec:ana} summarizes the performed analyses, including the description of the stellar and ionized gas decomposition performed by Pipe3D and the method adopted to detect and segregate the \hii\ regions (\pyHII, \citealp{lugosanchezespinosa2022}); Section \ref{sec:results} describes the main results of the current study, how the properties of \hii\ regions depend on global properties of their host galaxies (morphology and stellar mass), location within them (galactocentric distance) and the local properties of the underlying stellar population (luminosity-weighted age and metallicity). In addition we describe the main properties of the diffuse ionized gas, and the differences found in galaxies hosting and not hosting \hii\ regions; Section \ref{sec:discu} includes a discussion on the main results, with the final conclusions included in Section \ref{sec:conclu}.

Throughout the manuscript we adopt a standard $\Lambda$CDM cosmology with the following cosmological constants: $H_0$=71 km/s/Mpc, $\Omega_M$=0.27, $\Omega_\Lambda$=0.73. 

\section{Data and sample selection}
\label{sec:sample}

MUSE is an Integral Field Spectroscopy instrument that provides (i) until a seeing-limited spatial resolution (when we do not use the adaptive optic mode, as in our case), (ii) a Field-of-View (FoV) of 1'$\times$1' that corresponds to an inscribed circle of $\sim$30", (iii) a spatial sampling of 0.2"$\times$0.2" per spaxel, (iv) a spectral range from $\lambda$4750$\AA$ to $\lambda$9300$\AA$, (v) a spectral sampling of 1.25$\AA$, and (vi) a spectral resolution characterized by a FWHM that changes with the wavelength (being $\sim$2.4$\AA$ at the red part of the spectrum $\sim$7500$\AA$). These characteristics were defined by design to study objects at high-redshift \citep{muse}. However, they are particularly useful for the detailed exploration of nearby galaxies, e.g., GAs Stripping Phenomena in galaxies survey (GASP, \citealp{poggiantimorettigullieuszik2017}); The MUSE Atlas of Disks survey (MAD, \citealp{errozferrercarollodenbrok2019}) and Physics at High Angular resolution in Nearby GalaxieS survey (PHANGS-MUSE, \citealp{leroyhughesschruba2016}). 

\begin{figure}
    \minipage{0.46\textwidth}
    \includegraphics[width=\linewidth]
    {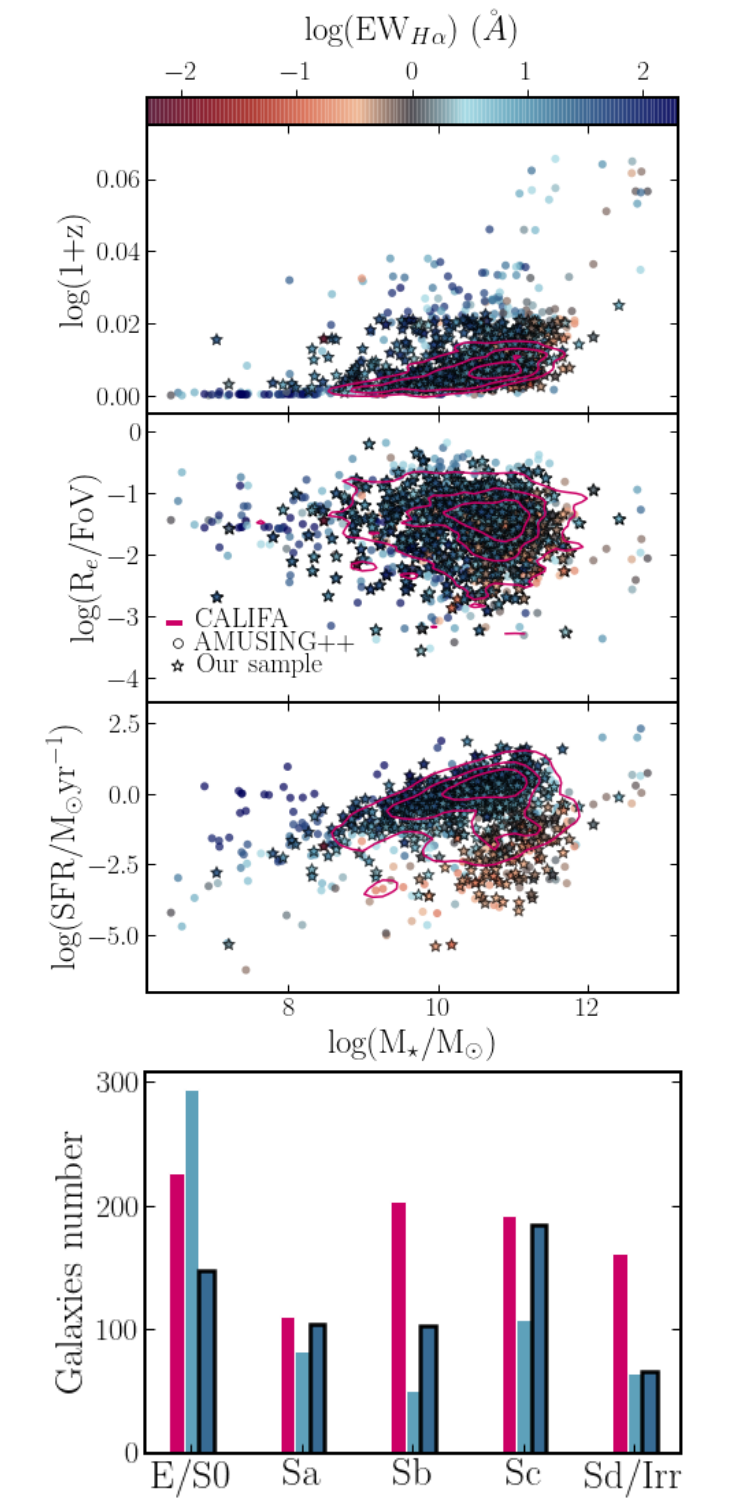}
    \endminipage
\caption{Comparison of the distributions of (i) the updated AMUSING++ sample (circles and blue bars without contour, 1053 galaxies), (ii) the final sub-sample selected for the current exploration (stars and blue bars with black contour, 678 galaxies), and (iii) the CALIFA sample (pink density contours and pink bars, 940 galaxies), archetype of a diameter selected sample, across the redshift ({\it top panel}), effective radius ({\it top middle panel}) and SFR ({\it bottom middle panel}) as a function of the stellar mass ({\it in the first three panels of top-down}). The color code represents the average EW(\Ha) across the FoV of the IFS data. The contours enclose a 90\%, 60\% and 30\% respectively of the CALIFA sample. In addition we show an histogram of the distribution of galaxies by morphology in the three samples ({\it bottom panel}).}
    \label{fig:1_showsample}
\end{figure}

We make use of these capabilities to study the spectroscopic properties of \hii\ regions in the nearby Universe. For doing so, we use the AMUSING++ compilation \citep{lopezcobasanchez2020}. This is an extension of the All-weather MUse Supernova Integral field Nearby Galaxies \citep[AMUSING,][]{amusing, amusing2} a survey to explore host galaxies of supernovae (SN-hosts) using MUSE data. The current compilation comprises all AMUSING targets plus additional archive observations, restricted to a similar redshift footprint ($z<$0.1) and impossing an  ``{\it a posteriori}" diameter selection. Diameter selections with galaxy size matching the FoV of the IFU are frequently adopted in IFS surveys to maximize the coverage and sampling of the optical extension of galaxies e.g., Calar Alto Legacy Integral Field Area survey \citep[CALIFA,][]{califapresentation}. 

The initial AMUSING++ compilation comprises 635 galaxies \citep{amusing, lopezcobasanchez2020}. Since the compilation was first presented we have continued exploring the ESO archive\footnote{\url{https://archive.eso.org/wdb/wdb/adp/phase3_main/form}}, downloading new observations that fulfill our basic criteria ($z<$0.1, diameter roughly matching the MUSE FoV). All these data are fully reduced by the version 0.18.5 of the MUSE pipeline \citep{weilbacherrothrousset2006, weilbacherstreicherurrutia2014}. So far, we have collected more than 1000 individual datacubes corresponding to a similar number of galaxies (1053). This corresponds to the updated AMUSING++ sample, after scanning through the ESO archive for new galaxies observed with MUSE following a similar criteria as those applied by \citealp{lopezcobasanchez2020} (original AMUSING++
sample). 

From this compilation we selected a sub-sample better suited for the purpose of the current exploration: (i) we limit the redshift range to 0.004 $<z<$ 0.06, ensuring a physical resolution in a range between $\sim$100 pc to $\sim$1 kpc; (ii) we limit the effective radii between 15$\arcsec$ and 120$\arcsec$ (therefore the MUSE FoV corresponds to 0.5-2 effective radius (R$_{e}$) for all galaxies). In addition, (i) we exclude galaxies for which the distance is uncertain (i.e., those  distant galaxies where the Hubble flow dominates above the intrinsic velocity of galaxies); (ii) we avoid galaxies strongly affected by the cosmic variance ($<$50 Mpc); hence we guarantee a resolution good enough to distinguish galaxy sub-structures such as bars \citep{lopezsanchezlin2022}, spiral arms \citep{sanchezsanchezperez2020}, and in particular good enough to explore giant \hii\ regions; and (iii) we perform a visual inspection cube by cube to rule out those observations taken under poor seeing conditions. Finally we search the morphological classification of all AMUSING++ galaxies in Hyperleda, NED and SIMBAD databases; only $\sim$85\% of galaxies have a morphological classification in any of these databases. The rest, we classify by eye. Our final sample comprises a set of 678 galaxies. 

Figure \ref{fig:1_showsample} shows the distribution of redshifts, effective radii and star-formation rates as a function of the stellar mass for the full compilation of MUSE data (1053 galaxies) and our sub-sample selected by redshift and effective radius (678 galaxies). For comparison purposes we include the distribution of galaxies of a well defined diameter selected sample \citep[CALIFA survey,][]{califapresentation}. We include, color coded, the average \Ha\ equivalent width, that traces the star-formation stage of each galaxy \citep{campssanchezlacerda2021,sanchezavilarodriguez2019}, and the distribution of morphological types for the three samples. 

If we consider the CALIFA sample as a archetypical diameter selected sub-sample of the galaxies in the local Universe, then comparison to our current sample will illustrate how representative our sample is. Indeed, in all panels we see that our sub-sample match roughtly well with the distribution traced by the CALIFA galaxies. In all figures $\sim$85\% of our galaxies is enclosed by the contour tracing 90\% of the CALIFA sample. Regarding the distribution by morphology, both samples cover the same set of galaxy types. However, we find some differences. In particular in the fraction of Sb and Sd/Irr galaxies that has a 50\% lower number in our sample than in CALIFA. It is known that this sample presents an excess of this kind of galaxies that is seen in the current comparison (\citealp{garciabenitozibettisanchez2015}). Finally we acknowledge a possible deficit of late-type galaxies in our sample (mainly Sd/Irr). Beside there is no significant differences in the remaining fraction of morphological types: E/S0, Sa, and Sc.
 
In summary, within the limitations inherent to a compilation and an {\it a posteriori} selection, the adopted AMUSING++ sub-sample roughly behaves as a diameter selected sample, covering a wide range of stellar masses, star-formation stages and morphological types in the considered wavelength range.  

\section{Analysis}
\label{sec:ana}

We analyze our sub-sample of 678 AMUSING++ datacubes using the \pipe\ pipeline to derive the main properties of the ionized emission lines and the stellar population components. Then, we use \pyHII\ to detect candidate \hii\ regions and decontaminate their properties from the possible contribution of the diffuse ionized gas (DIG). 

\subsection{\pipe\ analysis}
\label{sec:pipe3D}

\pipe\ is a pipeline developed to analyze IFS data. Its core package is FIT3D \citep{sanchezperezsanchez2016fit3d}. Both packages were originally written in Perl, although there is an update written in Python3 \citep{lacerdasanchezmejia2022}. \pipe\ is a popular tool used to analyze large data provide by IFS galaxy surveys such as CALIFA \citep{sanchezperezsanchez2016}, MaNGA \citep{sanchezbarreralacerda2022}, SAMI \citep{sanchezbarreralopez2019}, or AMUSING \citep{sanchezsanchezkawata2016}.

The main steps of the analysis performed by \pipe\ are: (i) a spatial binning in the continuum (V-band) to increase its signal-to-noise ratio (S/N$\sim$50); (ii) all the spectra in each bin are co-added generating a single spectrum per bin. It should be noted that during these first two steps the original shape of the spatial light distribution is preserved as much as possible (this is achieved by not co-adding spaxels with relative flux intensities with a difference larger than a 15\% between them), with the goal of limiting the loss of spatial information and obtaining precise and reliable stellar properties. (iii) Once the S/N ratio is increased, the stellar populations for each binned spectrum are fitted using a two step procedure; (iv) first, the velocity dispersion, stellar velocity and stellar dust attenuation are derived using a limited set of single stellar populations (SSPs) taken from stellar MIUSCAT templates \citep{vazdekisricciardellicenarro2012}; (v) second, we linearly fit a larger set of SSPs using the kinematics and the dust attenuation derived from the previous procedure and the GSD156 stellar population library \citep{fernandesperezgarcia2013}. GSD156 comprises a set of single stellar populations covering 39 ages and 4 metallicities (Z$_{\odot}$) with ranges from 1 Myr to 14 Gyr and from 0.2 to 1.6 Z$_{\odot}$, respectively. This library has been been widely used in the literature \citep{gonzalezcidgarcia2014,canosanchezzibetti2016, ibarrasanchezavila2016, mendezsanchezdelorenzo2019}. The advantages of this two step procedure are widely discussed in \cite{sanchezperezsanchez2016fit3d} and \cite{lacerdasanchezmejia2022}, but basically this helps to limit the degeneration between the velocity dispersion, metallicity, and dust attenuation. In the second step we perform a Monte Carlo iteration perturbing the original spectra using their corresponding errors, to propagate the errors through to those on the stellar parameters. From this second step we obtain the best model of the stellar continuum in each spaxel by re-scaling the model within each bin to the flux intensity at the corresponding spaxel. (vi) 
The properties of the stellar populations derived for each bin are propagated to each spaxel, obtaining 2D maps. These maps are packaged as slices for two different datacubes: SSP cube (average stellar properties) and SFH cube (the weights of the decomposition of the stellar population). (vii) Then, the best model spectra for the continuum is subtracted from original cube to create a gas-pure datacube. This cube contains the spectra of the ionized emission lines, noise and stellar population fitting residuals. (viii) The gas-pure cube is used to derive the properties of 30 emission lines by applying a non-parametric method of moment analysis \cite[see section 3.6 of][]{sanchezperezsanchez2016fit3d}. The complete list of emission lines was already published in Table C1 of \citet{lopezcobasanchez2020}. (ix) Then, we recover the main properties of ionized gas including the integrated flux intensities, line velocities, velocities dispersion and equivalent widths (together with the respective error on each property). Each map is packaged as an individual slice for a datacube called FLUX\_ELINES. (x) In addition, \pipe\ derives the spatial distribution of classical stellar indices. Those are determined from the measurement of a set of absorption line strength indices, using the stellar spectrum once subtracted the best model of the emission lines \cite[the Lick/IDS index system,][]{bursteinfabergaskell1984, faberfrielburstein1985}. Then, the code derives the equivalent width of each stellar index and a Monte Carlo iteration is performed to estimate the corresponding errors \cite[see Section 4.1.3 of][]{lacerdasanchezmejia2022}. The implementation of \pipe\ adopted for the current analysis estimates, the following stellar indices: H$\delta$, H$\delta$mod, H$\gamma$, H$\beta$, Mg$b$, Fe5270, Fe5335, and D4000. We should note that due to the limited wavelength range of the MUSE spectra, D4000, H$_{\delta}$, H$_{\delta}mod$ and H$_{\gamma}$ are not covered in the observed wavelength range and therefore their values correspond to fillers (i.e., they must be masked out).  

In summary, the analysis provides a set of 2D maps of each derived physical quantity or parameter, that are rearranged in a set of datacubes (with each channel corresponding to one property). We will refer to those cubes as dataproducts. There are four of them: SSP (comprising the average properties of the stellar populations, e.g., the Luminosity-Weighted (LW) age and metallicities), SFH (comprising the weights of each SSP for the best fitted model), FLUX\_ELINES (including the parameters of the emission lines, including fluxes, equivalent widths and velocities), and INDICES (including the values of a set of stellar Lick-Indices). This datamodel has been previous adopted and described in other articles e.g., \citet{ sanchezbarreralacerda2022}\footnote{\url{https://data.sdss.org/datamodel/files/MANGA_PIPE3D/MANGADRP_VER/PIPE3D_VER/PLATE/manga.Pipe3D.cube.html}}. It is important to note that throughout all the analysis, the same World Coordinate System (WCS) of the original cube is preserved and stored. For more details about the analysis performed by \pipe\ see \citet{sanchezperezsanchez2016fit3d} and \citet{lacerdasanchezmejia2022}.

As part of the post-processing of \pipe\ a set of characteristics properties for each galaxy is derived (\citealp{sanchez2020}) including the integrated star formation rate and stellar mass, as already shown in Fig. \ref{fig:1_showsample}. 

\subsection{\pyHII\ analysis}
\label{sec:pyHII}

\pyHII\ is a open access code written in Python, aimed to detect and extract the spectroscopic properties of \hii\ regions using high-spatial resolution IFS data such as the one provided by the AMUSING++ compilation. 
It essentially updates HIIexplorer and pyHIIexplorer algorithms (see \citealp{sanchezrosalesmarino2012} and \citealp{espi20}\footnote{\url{https://github.com/cespinosa/pyHIIexplorerV2}}), respectively). 

\begin{figure*}
    \minipage{0.99\textwidth}
    \includegraphics[width=\linewidth]
    {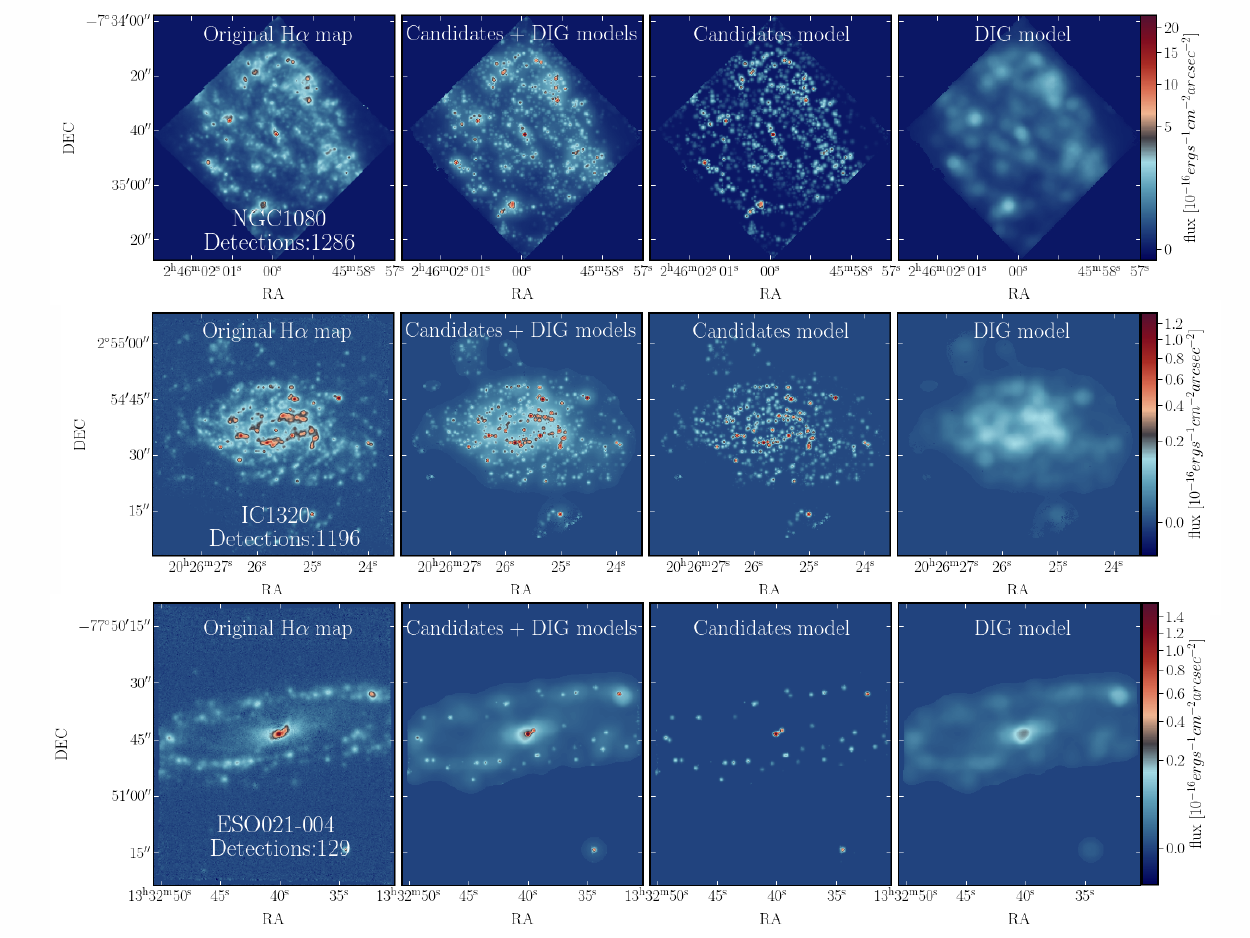}
    \endminipage
    \caption{Three examples of the use of \pyHII\ on three galaxies of different morphological type: NGC1084 (Sc, {\it top panels}), IC1320 (Sb, {\it middle panels}) and ESO021-004:S0 (S0, {\it bottom panels}). Each column in each row shows the \Ha\ intensity map corresponding to (i) the observational data derived by \pipe\ ({\it 1st panel}), (ii) the best combined model including the candidate \hii\ regions and the DIG ({\it 2nd panel}), (iii) the candidate \hii\ regions ({\it 3rd panel}) and (iv) the DIG model alone ({\it 4th panel}).}
        \label{fig:2}
\end{figure*}

\pyHII\ is divided into two main blocks: (i) detection of candidate \hii\ regions and (ii) extraction of the properties of the ionized gas emission lines and the underlying stellar populations, for each candidate. The main criterion for identifying an ionized region is that it must be a clumpy structure, centralized, peaked and with a relevant contrast with respect to the background (assumed to be a smooth ionization, i.e., diffuse ionized gas or DIG, at the considered resolution) in an emission line image. The major improvements with respect to its direct predecessor, HIIexplorer, is that it generates a model of the candidates and the DIG for each extracted spectroscopic property, decontaminating the former by the later. 

The first block comprises the routines that (i) detects the candidate \hii\ regions in an emission line map (\Ha\ map for our particular dataset), and (ii) generates a model of those regions and the DIG, segregating them. For detection, a modified version of the BLOB\_LOG algorithm included in the {\sc scikit-image} package\footnote{\url{https://scikit-image.org/}} is adopted. For the modelling, a Gaussian distribution for each candidate is assumed, while the model of the DIG results from a two step procedure involving the points with lowest contamination from the candidates and a kernel interpolation of the residual (once removed the candidates model).

The second block of routines included in \pyHII\ are focused on the extraction of the spectroscopic properties of the detected blobs. Currently, this block is optimized to use the \pipe\ dataproducts (described above). However, they can be easily adopted to be used with other IFS analysis and dataproducts, such as the one provided by pyCASSO \citep{deamorimgarciacid2017} or the MaNGA DAP \citep{belfiorewestfallschaefer2019, westfallcapellaribershady2019}.

For each physical or observational quantity provided by \pipe\ (e.g., any emission line flux intensity, the LW age of the underlying stellar population, or just a certain Lick index), the code extracts the corresponding value for each blob (already decontaminated by the DIG component, when suitable). In addition it provides two 2D distributions (models) of both (i) the blobs and (ii) the DIG, components. Finally, it provides errors of each extracted quantity and 2D map.

The extraction procedure returns to the user a set of tables and cubes comprising each extracted property. For the particular case of the dataproducts provided by \pipe\ there are four different tables for the individual blobs, and two set of datacubes (one for the blobs model and another for the DIG): (i) SSP, (ii) SFH, (iii) SSP, (iv) FLUX\_ELINES and (v) INDICES (described in Sec. \ref{sec:pipe3D}). 

During the development of \pyHII, we optimized the main input parameters of the code to be used on MUSE data, in particular, the parameter that defines the maximum size for the \hii\ region candidates. That parameter was fixed by default to the largest expected giant \hii\ region ($\sim$500pc), thus for our sample, we refine it galaxy by galaxy, using the value that will minimize the parameter of $\chi^{2}$ (for further details on the required parameters we refer to the GitHub\footnote{\url{https://github.com/sfsanchez72/pyHIIExplorer.git}}).

Figure \ref{fig:2} illustrates the detection procedure with \pyHII, showing three galaxies of different morphological type: NGC1084 (Sc), IC320 (Sb) and ESO021-004 (S0). We detected 1286, 1196 and 129 candidates for these late-, intermediate- and early-type spirals, respectively, in agreement with the expectations i.e., less \hii\ regions within earlier galaxy types.

We applied the above procedure to the full sample of 678 galaxies, obtaining a total of $\sim$150,000 candidate \hii\ regions, with their location in the sky and projected size. By construction a catalog of their a set of properties, including the emission line and stellar population ones provided by \pipe\ (as explained before) is obtained. Finally, maps are also obtained with the spatial distributions of the same properties for the DIG.

\subsection{HII regions selection}
\label{sec:selHII}

The selection of \hii\ regions from the candidates catalog described in the previous section is based on a set of criteria derived from previous studies (\citealp{kewleydopita2001, sanchezrosalesiglesias2014, morissetdelgadosanchez2016, lacerdacidcouto2018}). Thus we select as \hii\ region as: (i) a candidate with a minimum value of EW(\Ha)$>$6$\AA$, (ii) a value of S/N$>$3 in \Ha\ emission and finally (iii) be located on the [OIII]$\lambda$5007/\Hb\ (O3) versus [NII]$\lambda$6583/\Ha\ (N2) diagnostic diagram \citep{baldwin81} below the Kewley demarcation line \citep{kewleydopita2001}. Contrary to what was adopted in \cite{sanchezrosalesiglesias2014} and \cite{espi20}, we do not impose a criterion on the fraction of young stars contributing to the total luminosity to select \hii\ regions. The reason is that, in contrast to the IFS data used on those studies, the MUSE data do not cover the blue range of the optical spectrum below $\lambda$4400\AA, and thus misses those wavelengths more sensitive to the presence of young stars.

\begin{figure*}
    \minipage{1\textwidth}
    \includegraphics[width=\linewidth]
    {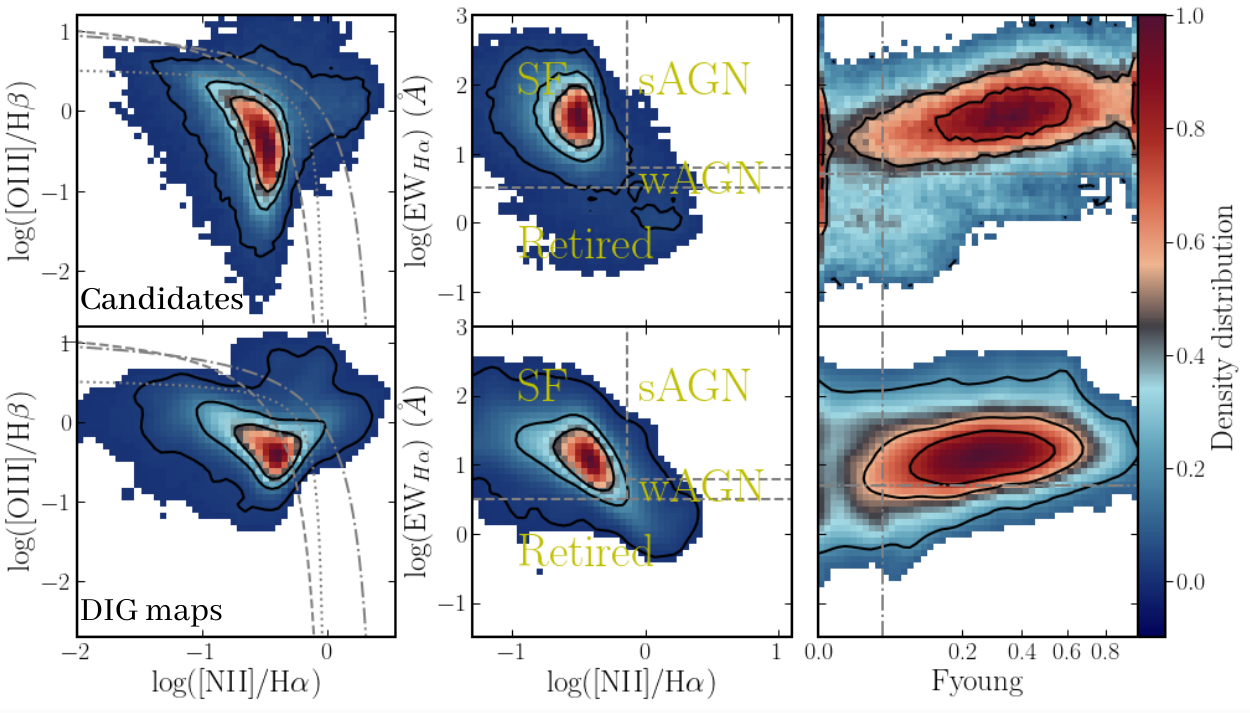}
    \endminipage
    \caption{Distribution of properties of the \hii\ regions candidates ({\it top panels}) and the DIG ({\it bottom panels}) across (i) the classical \oiii/\Hb\ vs. \nii/\Ha\ BPT diagnostic diagram \citep{baldwin81}, {\it left-panel}, (ii) the WHAN diagram \citet{cid11}, {\it central panels}, and (iii) the EW(\Ha) vs. fraction of young stars (WHAFY) diagram \citep{espi20}, {\it right-panel}. The colors and contours in each diagram represented the relative density of the distribution, with each successive contour encircling 90\%, 50\% and 15\% of the data points. The different grey lines in each diagram corresponds to different boundaries/demarcations lines adopted to distinguish between different ionizing sources: (i) in the BPT diagram, the gray dashed, dot-dashed and dotted lines correspond to the \citet{kauffmanheckmantremonti2003}, \citet{kewleydopita2001} and \citet{espi20}, demarcation lines respectively; (ii) in the WHAN diagram, the boundaries indicate the location of SF, strong AGN, weak AGN and retired ionized regions (or retired SF regions, e.g., \citealp{canosanchezzibetti2016}); and (iii) in the WHAFY diagram the vertical line corresponds to the previously adopted boundary between retired and SF regions \citep{espi20}.}
    \label{fig:3_diagrams_td}
\end{figure*}

Figure \ref{fig:3_diagrams_td} shows three different diagnostic diagrams built with the information derived from the \pyHII\ analysis for the catalog of \hii\ region candidates and the DIG maps. From the upper panels, corresponding to the candidates selected by \pyHII, we highlight that only by recovering clumpy structures with strong \Ha\ emission we do recover those mostly located in the regions of the diagram usually associated to SF ionization. This is more clear in the BPT and EW(\Ha) versus \nii/\Ha\ (WHAN) diagrams (upper left and upper middle panels respectively ), where the position of most of the contours that enclose the density of points are in the area of star formation, except for the contour that encloses 90\% in both diagrams. Thus, only a small fraction of the clumpy ionized regions selected by \pyHII\ are indeed located in the composite and LINER areas of the BPT and retired area of the WHAN diagram.

On the other hand, in the case of the lower panels corresponding to the diffuse gas of all galaxies, we notice that the regions covered by the density contours are wider than those of the \hii\ region candidates in the diagrams (BPT and WHAN). For example, in the BPT diagram, there is a greater coverage of the diffuse emission towards the AGN zones (principally the contour of 90\%), with a large fraction of spaxels covering the intermediate region between the Kauffman and Kewley demarcation lines (\citealp{kauffmanheckmantremonti2003}, and \citealp{kewleydopita2001}, respectively). Something similar happens in the case of the WHAN diagram where the distribution of the DIG reaches areas usually assigned to retired regions, weak- and strong- AGNs.  

Finally, we explore the diagram of EW(\Ha) as a function of Fyoung, where Fyoung, is the fraction of light in the V-band with ages $<$350Myr of the underlying stellar population, or WHAFY diagram hereafter. We observe that in this diagram both the clumpy ionized regions and the DIG are located preferentially above the EW(\Ha)=$>$6\AA\ and Fyoung$>$0.4\%. The latter was the limit adopted by \cite{espi20} to distinguish between \hii\ regions and other ionized sources. As we notice already it would not be useful for our particular analysis due to the limited wavelenght range. Thus, we do not adopt any cut based on the WHAFY diagram in our selection of \hii\ regions.  

Once the criteria described above are applied to the candidates, we recover 52,371 \hii\ regions in 539 different galaxies, each \hii\ region must meet the above 3 criteria (EW(\Ha)$>$6\AA, a minimum value of S/N, and the location below Kewley line). For the remaining 139 galaxies, we do not recover any \hii\ region. 

\section{Results}
\label{sec:results}

\subsection{\hii\ regions along the Hubble sequence and stellar mass}
\label{sec:hubbleseq}

As part of the exploration of the final catalogue, we segregate the \hii\ regions according to: (i) properties of the host-galaxy such as morphology and integrated stellar mass, and (ii) intrinsic properties of the \hii\ regions such as its galactocentric distance, average value of the luminosity-weighted age and metallicity of the underlying stellar population.

\begin{figure*}
    \minipage{0.99\textwidth}
    \includegraphics[width=\linewidth]
    {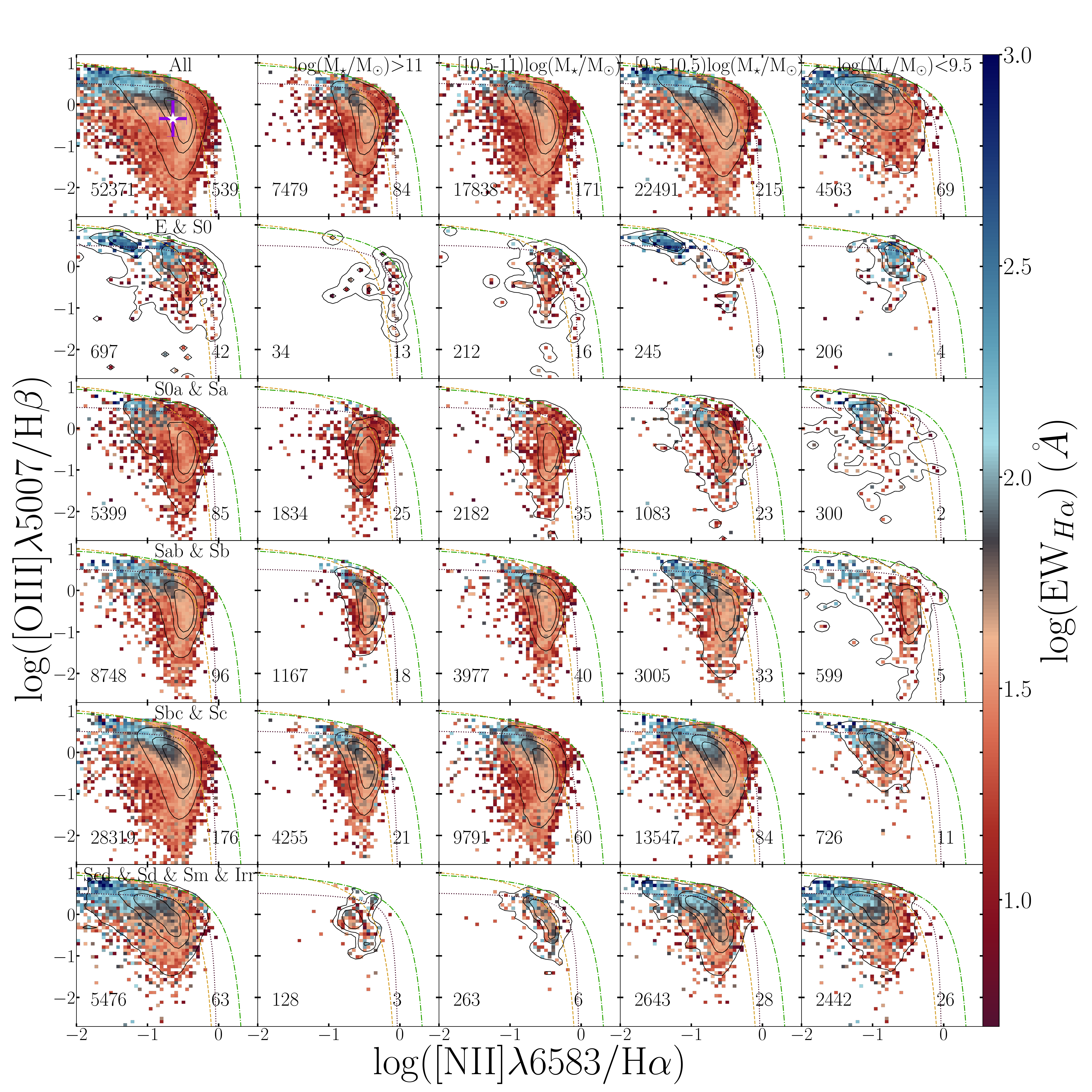}
    \endminipage
    \caption{Same \oiii/\Hb\ vs. \nii/\Ha\ BPT diagnostic diagram shown in left panels of the Fig. \ref{fig:3_diagrams_td} in each panel for our final sample of \hii\ regions segregated by the morphology and stellar mass of their respective host galaxies. From top to bottom (left to right) galaxies are grouped by morphology (stellar mass), with each subgroup indicated in the legend in the top of each panel. The number of regions is indicated in the bottom-left panel and the number of galaxies corresponding to that bin in the bottom-right panel. The color bar indicates the average value of the EW(\Ha) of the \hii\ regions at the corresponding location in the BPT diagram. The white star is the mean value of the catalog, and the purple lines are the typical/averaged errors. In each panel, yellow dashed, blue dot-dashed and black-dotted lines represent \citet{kauffmanheckmantremonti2003}, \citet{kewleydopita2001} and \citet{espi20} demarcation lines, respectively. Contours correspond to the density distribution, enclosing 90\%, 50\% and 15\% of the regions, respectively.}
    \label{fig:4_hii}
\end{figure*}

First, we explore in figure \ref{fig:4_hii} the location of the \hii\ regions across the BPT diagram segregated by the morphology and stellar mass of the host-galaxy. The upper-left panel comprises the full \hii\ region catalog, independent of the properties of the host galaxies. Each column comprises galaxies of different mass range, from more massive (2nd column) to less massive (5th column). On the other hand, each row corresponds to galaxies with different morphology, from earlier types (2nd row) to later types (6th row). This way the intersection between each row and column corresponds to galaxies of a certain morphology (row) and within a certain stellar mass range (column).

\begin{figure*}
    \minipage{0.99\textwidth}
    \includegraphics[width=\linewidth]
    {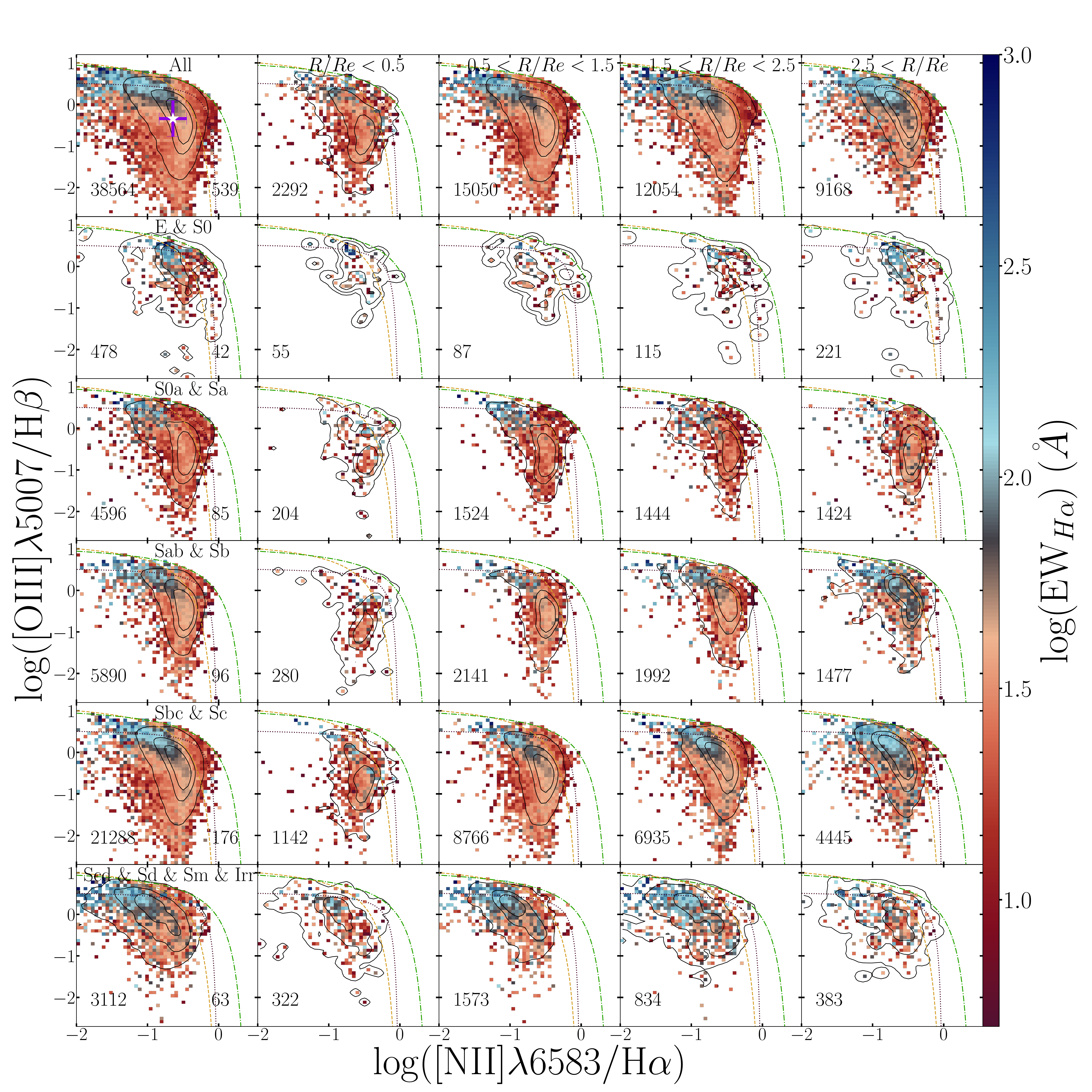}
    \endminipage
    \caption{Similar figure as the one shown in Fig. \ref{fig:4_hii}, using the same sample and nomenclature. In this case each row corresponds to the same morphological types described in that figure. However, each column corresponds to \hii\ regions in different galactocentric distances normalized to the effective radius ($Re$). Each distance is indicated in the top legend. Colour bar, white star, purple lines, demarcation lines and density contours are same as Fig. \ref{fig:4_hii}.}
    \label{fig:5_radii}
\end{figure*}

In quantitative terms, the highest percentage of \hii\ regions is present in intermediate-type galaxies (Sbc \& Sc: 54\%). In the particular case of galaxies types Scd, Sd, Sm, and Irr, there is only $\sim$10.4\% of the \hii\ regions. This is due to the low number of such galaxies with respect to all those that have morphological classification. As we know massive galaxies generally correspond to earlier types \citep{labbevandokkumnelson2022}, so it is expected that the number of \hii\ regions is smaller in comparison to those in intermediate and less massive galaxies (or late-type, e.g., \citealp{kennicutt1989}). As the stellar mass increases, the percentage of \hii\ regions decreases ([9.5-10.5]log(M$_{*}$/M$_{\odot}$): 42.9\%, [10.5-11]log(M$_{*}$/M$_{\odot}$): 34\% and log(M$_{*}$/M$_{\odot}$)$>$11: 14.2\%). Except for the lowest mass bin included in our sample, whose percentage is relatively low log(M$_{*}$/M$_{\odot}$)$<$9.5: 8.7\% due to the small number of such galaxies in the total sample (see histogram in Fig. \ref{fig:1_showsample}).  

Figure \ref{fig:4_hii} shows evident trends in the distribution of \hii\ regions in the BPT diagram for different masses and morphologies. In early-type massive galaxies they are mostly located in the right-end of the distribution, presenting in general low values of the EW(\Ha). On the other hand, for late-type/low-mass galaxies \hii\ regions are more frequently found in the left-end of the distribution with high EW(\Ha) values (this is valid for spiral galaxies in general). 

These trends are similar to those reported in previous works such as \cite{espi20} and \cite{sanchezperezrosalesortega2015}. E \& S0 galaxies deviate from this pattern since their \hii\ regions are located in the upper left-end of the distribution with high EW(\Ha) values, without following a clear/obvious distribution. We will address this particular case in the discussion. In general all the intermediate panels describe to a greater or lesser extent these general trends according to what the statistic allows. 

\subsection{\hii\ regions along galactocentric distances}
\label{sec:galdistan}

Continuing the exploration of the final catalogue of the \hii\ regions, we also segregated it according to its \hii\ region distance from the center of each host galaxy. To do this, we derive the galactocentric distance of each \hii\ region using its coordinates (as well as the coordinates of the centers of each galaxy), deprojecting them using the mean inclination and position angle of each galaxy and normalizing those distances to the effective radius of each galaxy.

Figure \ref{fig:5_radii} shows the distribution of our final catalogue of \hii\ regions across the BPT diagram separated by their galactocentric distances and morphology of the host-galaxy. Left to right panels comprise the regions in the (i) the center of the galaxies roughly coincident with the bulge ($R/Re<0.5$); (ii) the transition region between the bulge and the disk ($0.5<R/Re<1.5$); (iii) the disk ($1.5<R/Re<2.5$) and (iv) the outer disk regions ($2.5<R/Re$). From top to bottom the panels comprise the segregation by morphology in the same way as in Fig. \ref{fig:4_hii}. The color indicates the mean value of EW(\Ha). For this particular segregation, the highest percentages of \hii\ regions are found at the intermediate disk ($\sim$39\%) and for the Sbc morphological type ($\sim$55\%).

\begin{figure*}
    \minipage{0.99\textwidth}
    \includegraphics[width=\linewidth]
    {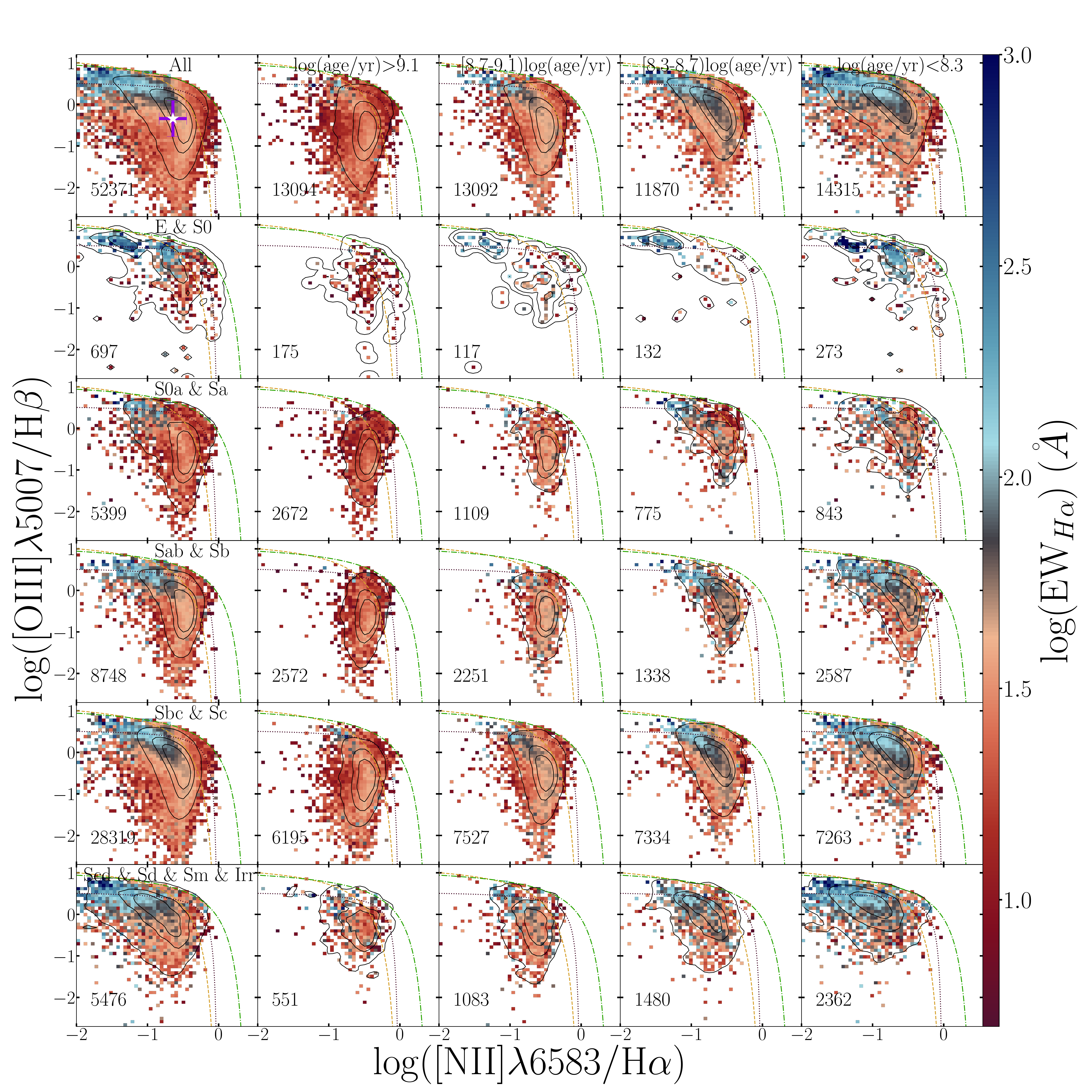}
    \endminipage
    \caption{Similar figure as the one shown in Fig. \ref{fig:4_hii}, using the same sample and nomenclature. In this case each row corresponds to the same morphological types described in that figure. However, each column corresponds to \hii\ regions with different average values of the luminosity-weighted age of the underlying stellar population. Each age range shown in the top legend. Colour bar, white star, purple lines, demarcation lines and density contours are same as Fig. \ref{fig:4_hii}.}
    \label{fig:6_age}
\end{figure*}

We find similar trends as those observed in the Fig. \ref{fig:4_hii} in agreement with the expectations: the \hii\ regions located in the central areas of their host galaxy are mainly located in the right-end area of the BPT diagram with low values of EW(\Ha). On the other hand, the distributions of the \hii\ regions further away from the center of their respective host galaxies, are more frequently located in the upper-left region of the BPT diagram, with higher values of EW(\Ha). This is in agreement with the results of previous studies such as \cite{kennicuttkeelblaha1989} and \cite{hofilippenkosargent1997}, that reported that the \hii\ regions in the central regions of galaxies present higher values of \nii\ than those in outer areas. More recent explorations, such as \cite{espi20} also found a radial trend in the properties of \hii\ regions that follows a similar pattern as the one observed here. On the contrary, other authors did not find such clear radial trends at those found here, e.g., \cite{veilleuxosterbrockdonald1987}.

\begin{figure*}
    \minipage{0.99\textwidth}
    \includegraphics[width=\linewidth]
    {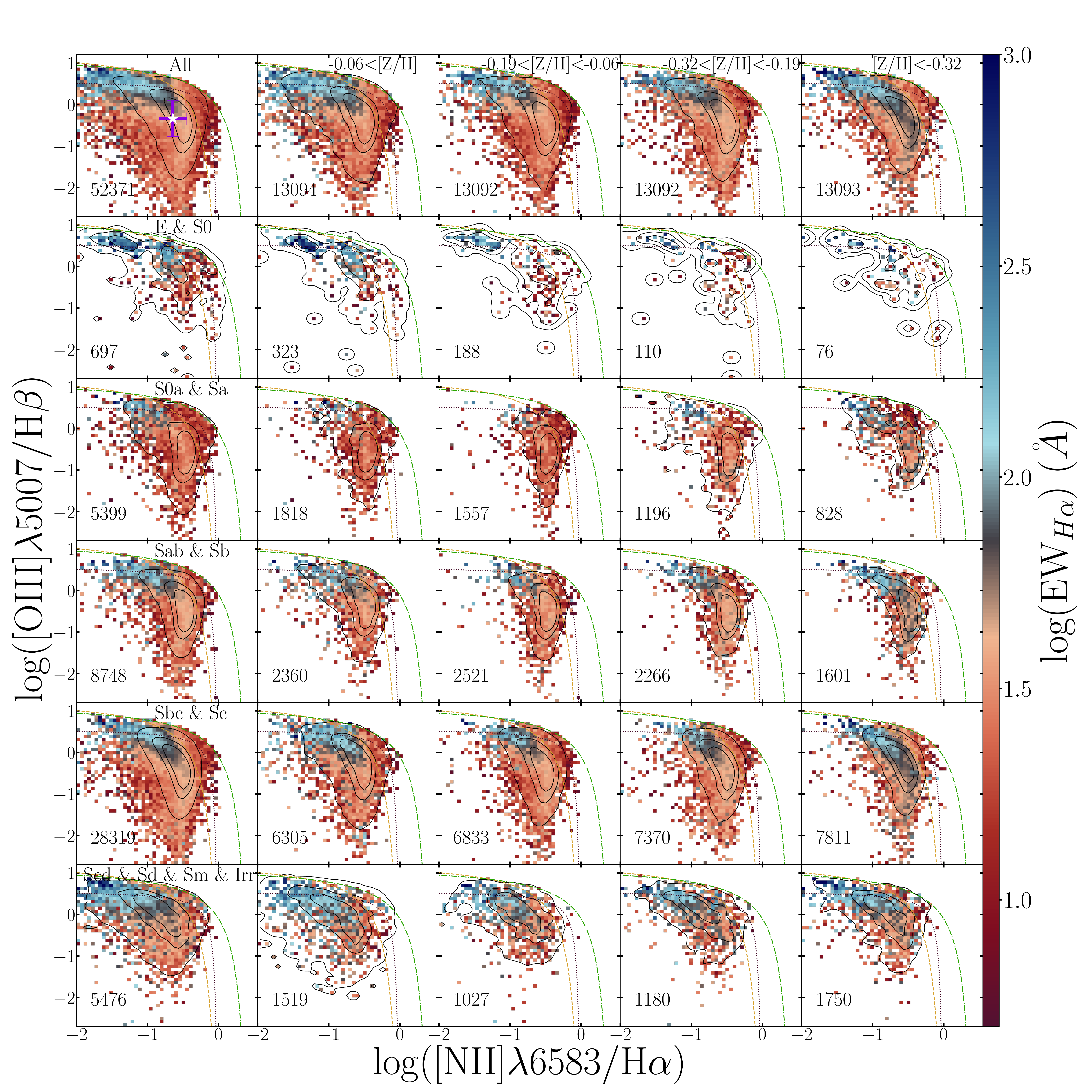}
    \endminipage
    \caption{Similar figure as the one shown in Fig. \ref{fig:4_hii}, using the same sample and nomenclature. In this case each row corresponds to the same morphological types described in that figure, but with the columns corresponding to a segregation by the average value of the luminosity-weighted metallicity of the underlying stellar population. The units are [Z/H]=log(Z/Z$_{\odot}$), where Z$_{\odot}$=0.02. Colour bar, white star, purple lines, demarcation lines and density contours are same as Fig. \ref{fig:4_hii}.}
    \label{fig:7_met}
\end{figure*}

\subsection{\hii\ regions and the age of the underlying stellar population}
\label{sec:agepop}

In the two previous sections, we confirmed the results by \cite{sanchezperezrosalesortega2015} and \cite{espi20}, noting that the distribution of the \hii\ regions along the classical BPT diagram is directly related to the morphology of the host galaxy, its mass, and the galactocentric distance from the respective regions. We need to recall that in these previous studies and in this work, we analyzed all the stellar population at location of the \hii\ regions (see Sec. \ref{sec:pyHII}) without segregating the ionizing stellar component as in the works of \citet{smithnorriscrowther2002} and \citet{mirallesdiazrosales2014}. 

\begin{figure*}
    \minipage{0.99\textwidth}
    \includegraphics[width=\linewidth]
    {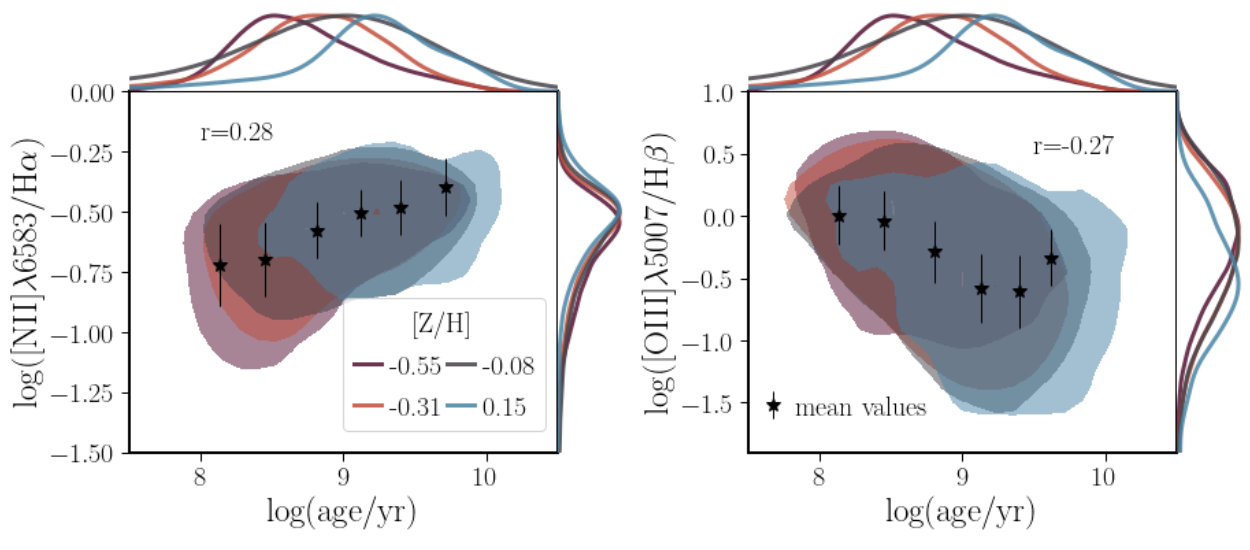}
    \endminipage
    \caption{Distributions of the \nii$\lambda$6583/\Ha\ ({\it left panel}) and \oiii$\lambda$5007/\Hb\ ({\it right panel}) line ratios as a function of the luminosity-weighted (LW) age for the final sample of \hii\ regions. For both panels each different color corresponds to the distributions segregated by the LW metallicity of the underlying stellar populations as indicated in the legend, with the average [Z/H] of each group being: -0.55 (red), -0.08 (orange), -0.31 (gray), and 0.15 (blue). Black stars correspond to the average values of the total distribution of points, with error bars that corresponding to the standard deviation.}
    \label{fig:10_age_zh}
\end{figure*}

Both studies also connected the properties of the underlying stellar population and the location of the \hii\ regions in the BPT diagram. The reported connection is based on the fact that the \hii\ regions located in the areas closer to the center of the late-type galaxies present similarities with those found in massive early-type galaxies, which is reflected in their location in the BPT diagram (right/lower-right), as well as the \hii\ regions at larger galactocentric distances having similarities with those located in less massive or late-type galaxies (left/upper left). Those two regimes, early-types/centers versus late-types/outer regions, do indeed present the same kind of behavior of stellar populations \citep{blantonmoustakas2009, gonzalezperezcid2014}. 

In Fig. \ref{fig:6_age} we show the distribution of our final catalog along the BPT diagram segregated by the average values of the luminosity-weighted age of the underlying stellar population and by morphology of the host-galaxy. From left-to-right the panels contain the segregation from the oldest to the youngest stellar populations, and from top-to-bottom, the morphological segregation already shown in Fig. \ref{fig:4_hii} and \ref{fig:5_radii}. Once more, the color indicates the mean value of the EW(\Ha) of all \hii\ regions in the same BPT loci. 

The largest fraction of \hii\ regions are now found in regions which underlying stellar population is less than log(age/yr)$<$8.3 (27.3\%) and are mostly found in Sbc and Sc galaxies. For the most part, the regions in which underlying stellar populations have the age  log(age/yr)$<$8.3 shows a tendency towards left/upper-left areas of the BPT diagram except for those found in the earliest type galaxies (E \& S0) that have high values of EW(\Ha) too. As for the regions with older underlying stellar populations, log(age/yr)$>$9.1 and [8.7-9.1)log(age/yr), have a trend of lower values of EW(\Ha) being located mainly in the right/lower-right areas in the respective BPT diagrams. 

\subsection{\hii\ regions and metallicity of the underlying stellar population}
\label{sec:metpop}

Figure \ref{fig:7_met} shows the distribution of our final catalogue of \hii\ regions along the BPT diagram segregated by the metallicity of the underlying stellar population and morphology of the host-galaxy. From left-to-right each panel contains the distribution for the regions with different stellar metallicity of the underlying stellar populations, from metal-rich to metal-poor. From top-to-bottom each panel shows the distributions separated by morphology, following the same scheme already adopted in Figs. \ref{fig:4_hii}-\ref{fig:6_age}. Once more, the EW(\Ha) is displayed by the color. In quantitative terms, in our catalogue the largest fraction of \hii\ regions corresponding to -0.32$>$[Z/H] of stellar metallicity present mainly in Sbc and Sc galaxies. 

While trends are not as clear as in the case of other parameters such as the age of the underlying stellar populations, in general those regions with metal-rich underlying stellar populations are located at the right/bottom end of the distribution (regardless of the morphological type of the host-galaxy), while that those with more metal-poor underlying stellar populations are found in the upper-left regime of the distribution. 

It is known that early-type galaxies are dominated by old, metal-rich stellar populations, while in late-type galaxies, the central regions have stellar populations similar to early-type galaxies and the outer and/or disk parts are dominates by younger and more metal poor stellar populations (e.g., \citealp{gonzalezcidgarcia2014, garciagonzalesperez2017, sanchez2020, sanchezwalcherlopez2021}). As the location in the BPT diagram of \hii\ regions follow a similar pattern, it has been suggested that the physical connection between both patterns is the properties of the underlying stellar populations (\citealp{sanchezperezrosalesortega2015, espi20}). In the next section we explore this particular scenario. 

\subsection{Dependence of O3$\equiv$\oiii/\Hb\ and N2$\equiv$\nii/\Ha\ line ratios with the properties of the underlying stellar populations}
\label{sec:conec}

As indicated in previous sections our results suggest a connection between the location in the BPT diagram of the \hii\ regions and their properties of underlying stellar populations. We explore this connection in a more quantitatively way in this section. Figure \ref{fig:10_age_zh} shows the distribution of N2 line ratio (left panel) and the O3 line ratio (right panel) both versus the luminosity-weighted age for the final catalogue of \hii\ regions. 

We include in both panels the average value of the line ratios (N2 or O3) in bins 0.3 dex in log(age). Those average values show the described trend in a more clear way and as the first approximation. We used the binned values to calculate the Pearson correlation coefficients of line ratios and age, obtaining the values that show both panels (r=0.28 and r=-0.27)

To deepen the characterization, we derive the Pearson correlation coefficients of each variables pair and fit each line ratios (N2 and O3) to stellar properties (age and [Z/H]), then we explore the dependency with the following stellar property after removing the first and finally, again we calculate the Pearson correlation coefficient. We use the complete distribution of \hii\ regions (without binning) to obtain the best correlation values. Table \ref{tab:1} summarizes the results of this analysis.  

The primary relation of the age (metallicity) of the underlying stellar population has a correlation coefficient of 0.37 (0.13) with the N2 line ratio and an anti-correlation of -0.37 (-0.17) with the O3 line ratio. On the other hand, the secondary relation of the age (metallicity) of the underlying stellar population has a correlation of 0.35 (0.005) with the residual of  N2 line ratio and an anti-correlation of -0.34 (-0.10) with the O3 line ratio. These values indicate a moderate (weak) dependence between the parameters.

We should note that the correlations are stronger when they are estimated with the complete distribution of \hii\ regions as shown in Table \ref{tab:1} than when are estimated with the binned values as exemplified in the Fig. \ref{fig:10_age_zh} and that the age of the underlying stellar populations has a stronger connection than stellar metallicity with the line ratios. 

\begin{table}
\begin{center}
\caption{Dependencies between line ratios (N2 and O3) with age and metallicity of underlying stellar populations. (i) Line ratio. (ii) Property of stellar populations to fit. (iii) Correlation coefficient between line ratio and the stellar property to fit. (iii) Property of residual stellar populations. (iv) Correlation coefficient between the residue (obtained of best model and the values of the original line ratio) and the property of residual stellar populations.}
\begin{tabular}{ |c|c|c|c|c| } 
\hline
Ind.&Par. vs&C. C. (orig.)&Res. vs&C. C. ($\Delta$Ind$_{model par.}$, res.)\\
\hline
N2 & Age & 0.37 & [Z/H] & 0.055 \\ 
N2 & [Z/H] & 0.13 & Age & 0.35\\ 
O3 & Age & -0.37 & [Z/H] & -0.10\\
O3 & [Z/H] & -0.17 & Age & -0.34\\
\hline
\label{tab:1}
\end{tabular}
\end{center}
\end{table}

\subsection{Different types of DIG}
\label{sec:dig_types}

As described in Sec. \ref{sec:pyHII} our analysis using \pyHII\ provides a two dimensional distribution of the diffuse ionized gas of each analyzed property for each galaxy.

According to previous works, DIG has been argued to be ionized by different sources such as: (i) heating due to cosmic rays \citep[e.g.,][]{reynoldscox1992}; (ii) X-rays \citep[e.g.,][]{domgorgenmathis1994} or (iii) dust grains \citep[e.g.,][]{draine1978}; (iv) turbulent dissipation \citep[e.g.,][]{binetteflores2009}; (v) photoionization by hot low-mass envolved stars (or HOLMES; e.g., \citealp{floresfajardomorissetstasinska2011}); (vi) fast shocks from supernova winds \citep[e.g.,][]{allengrovesdopita2008} or (vii) leakage of Lyman continuum photons from \hii\ regions with young stars (mainly type O and B; \citealp[e.g.,][]{weilbachermonrealiberoverhamme2018}). Thus, in a galaxy there may coexists ``different types of DIG" or it may be that one specific type dominates in a specific galaxy (or in a specific area). In this section we present our exploration of the properties of the DIG.  

First, we verify that the properties of our derived DIG are consistent with the patterns already described by previous studies. \citealp{lacerdacidcouto2018} describes the existence of at least two distinct types of DIG, a hot-phase (more frequently found in early-type galaxies), and a warm phase (more frequently found in late-type galaxies). \citealp{espi20} demonstrated that indeed there is a gradual transition between one and the other as later (or earlier) is the host galaxy. Hot-phase present lower values for the EW(\Ha) (1-3\AA), while warm phase present larger values (3-10\AA). We replicate those results in here by showing the distribution of EW(\Ha) of the DIG maps for four galaxies of different morphological types.

\begin{figure}
    \minipage{0.48\textwidth}
    \includegraphics[width=\linewidth]{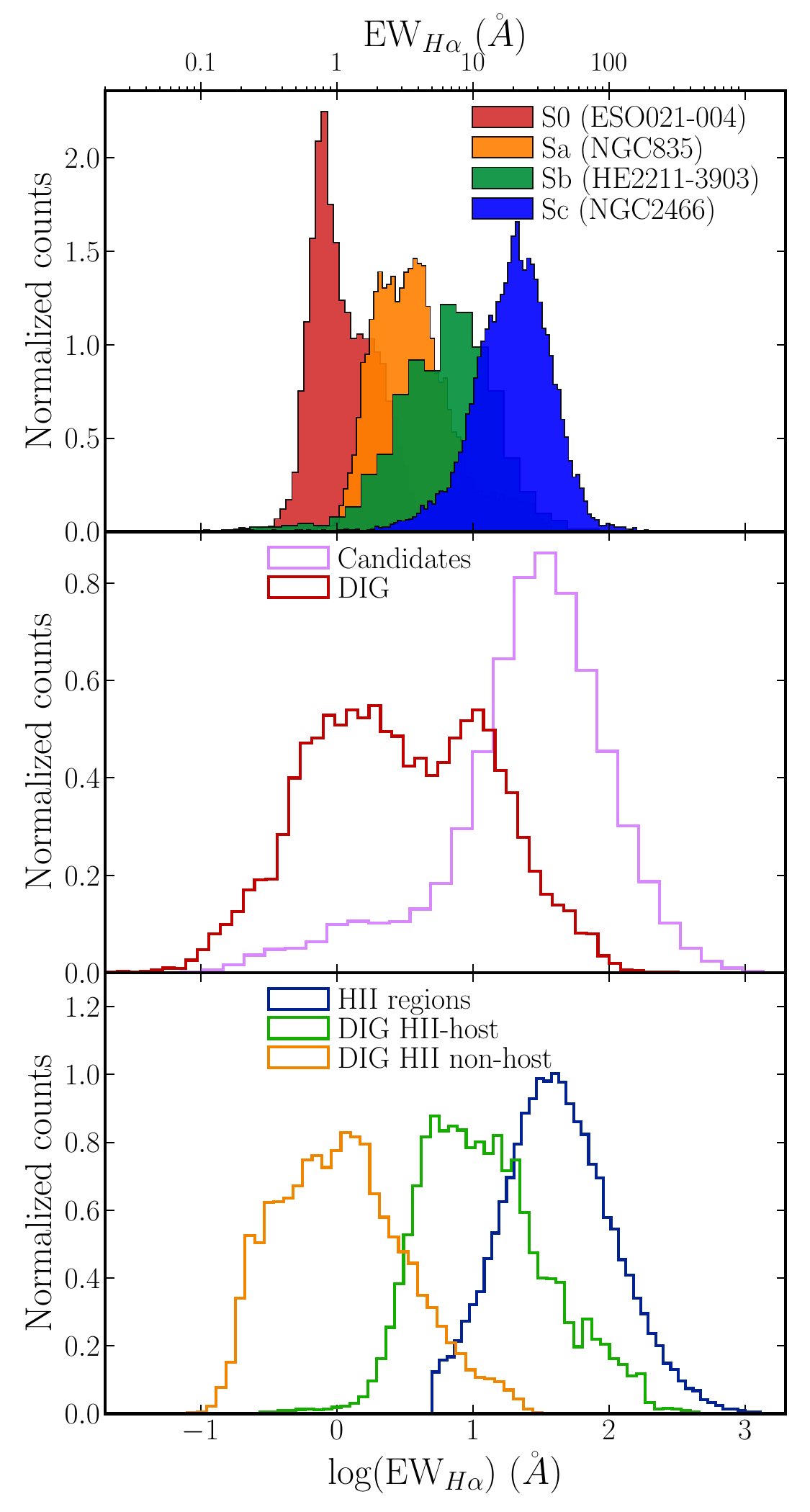}
    \endminipage
    \caption{Normalized histograms of the distribution of EW(\Ha) for the different components provided by our analysis: (i) corresponding to spaxels of DIG (spatially resolved and masked with segmentation maps) derived by \pyHII, excluding the regions corresponding to \Ha\ blobs (candidate \hii\ regions) for 4 different galaxies selected along the Hubble sequence ({\it top panel}), (ii) for the segregation of \hii\ regions candidates, and DIG spaxels ({\it middle panel}), (iii) for the \hii\ regions selected from the candidates, DIG of those galaxies hosting \hii\ regions and DIG of those galaxies not hosting \hii\ regions ({\it bottom panel}). Each bin shows the number of counts in the bin divided by the bin width and the total number of counts; thus the area under each histogram is integrated to 1.0.}
    \label{fig:8_dig}
\end{figure}

\begin{figure*}
    \minipage{0.99\textwidth}
    \includegraphics[width=\linewidth]
    {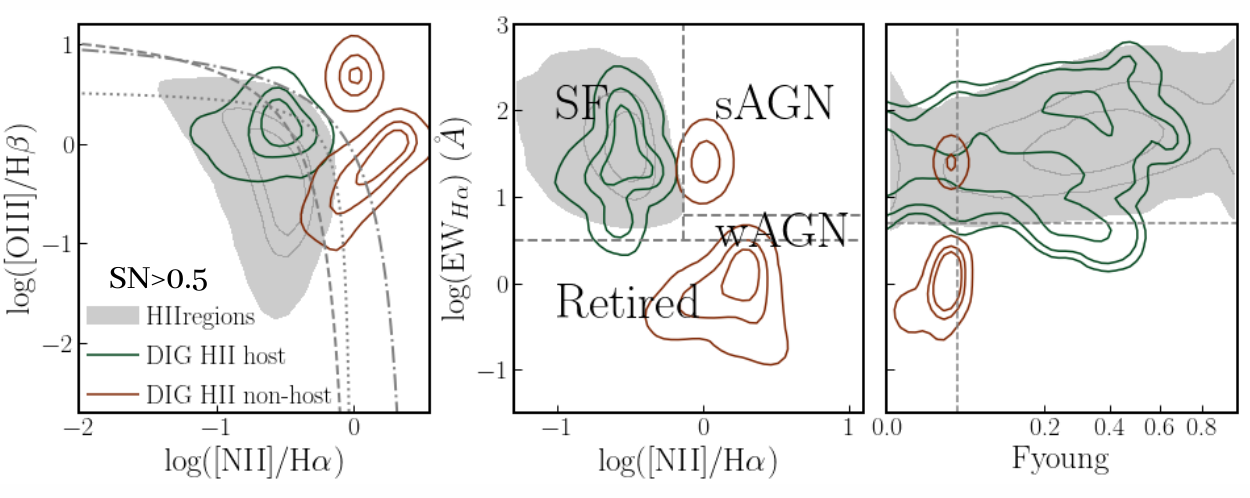}
    \endminipage
    \caption{Similar panels shown in Fig. \ref{fig:3_diagrams_td} for the three components shown in the bottom panel of Fig. \ref{fig:8_dig}: \hii\ regions (gray contours), and DIG maps for galaxies hosting (green contours) and not hosting of \hii\ regions (orange contours). Contours and demarcation/boundary lines are the same shown in panels of the Fig. \ref{fig:3_diagrams_td}.}
    \label{fig:9_diagrams}
\end{figure*}

Figure \ref{fig:8_dig} top panel, shows the histogram of the EW(\Ha) for the DIG maps derived for these archetype galaxies. As expected the earlier-type galaxy (S0: ESO021-004) has the lowest EW(\Ha) values, that in average increases from the earlier spirals (Sa: NGC835), towards the later ones (Sb: HE2211-3903), with the highest values found for the latest galaxy type (Sc: NGC2466). This trend agrees with previous results \citep{lacerdacidcouto2018, espi20}. On the other hand, in the middle panel, we show the histograms for the two components derived by \pyHII, the clumpy regions or blobs and the DIG maps. Both distributions cover different ranges, and it is possible to differentiate between them. The EW(\Ha) for the clumpy regions or blobs determinate by \pyHII\ present a well defined peak centered at $\sim$50\AA, with values covering a range up to $\sim$1000\AA, and a tail of very few points extending below 3\AA. On the contrary, the histogram for the DIG presents in general lower values of the EW, covering a range between 0.1\AA\ and 100\AA. The distribution of equivalent widths of the DIG is clearly bimodal, with two peaks, one centred at $\sim$1\AA\ and another at $\sim$10\AA, with a valley at $\sim$3-4\AA\ that divides the number of points in a 60/40\%, for low- and high-EWs respectively. This agrees with the results by \cite{belfioresantorogroves2022} that describes two main different types of diffuse emission in the PHANGS-MUSE sample with similar properties to those described here. Finally, in the bottom panel, we show a similar histogram for the DIG derived for (i) galaxies without any detected \hii\ region (DIG \hii\ not host) where the EW(\Ha) covers the lowest value range, clearly corresponding to the low-EW regime of the previous bimodal distribution (EW(\Ha)=0.1\AA\ and $\sim$EW(\Ha)=10\AA); (ii) galaxies with detected \hii\ regions (DIG \hii-host), where the EW(\Ha) cover the high-EW range of the previous bimodal distribution (EW(\Ha)=1\AA\ and EW(\Ha)=100\AA); and finally (iii) our sample of selected \hii\ regions where the EW values are clearly larger than any of the previous two (between EW(\Ha)=5\AA\ to EW(\Ha)=1000\AA), with a well defined peak and an almost symmetrical distribution (in the logarithm space). In this panel the nature of the bimodal distribution appreciated in the DIG shown in previous panel (i.e., non segregating between galaxies hosting or not \hii\ regions) is naturally understood. 

A possible interpretation to these result is that the dominant ionizing source of the DIG changes with the morphological type. As discussed in \cite{sanchez2020}, in early-type galaxies, with lower numbers of \hii\ regions and older stellar populations the most suitable candidate ionize the DIG is either HOLMES, post-AGB in general \citep{singhvandevenjahnke2013, belfioremaiolinomaraston2017} with a possible contribution of low-velocity shocks \citep{dopitasutherland1996}. On the other hand, in later type galaxies, with a larger coverage of \hii\ regions, photon {\it leaking} from those regions may produce a large fraction of the diffuse ionization \citep{beckmanrozaszurita2000, zuritabeckmanrozas2002}. Therefore, DIG near to \hii\ regions is most probably dominated by {\it leaking}, while that located at larger distances has a stronger contribution of ionization due to old stars (in agreement with \citealp{belfioresantorogroves2022}). 

\subsection{Location of the two DIGs in the diagnostic diagrams}
\label{sec:class_dig}

With the distinction between the DIG for galaxies hosting or not hosting of \hii\ regions described in the previous section, we decided to explore the location of those components in the classical diagnostic diagram. Figure \ref{fig:9_diagrams} shows the same three diagnostic diagrams shown in the Fig. \ref{fig:3_diagrams_td} (BPT diagram, WHAN diagram and the distribution of EW(\Ha) and WHAFY diagram) for the three components shown in the bottom panel of Fig. \ref{fig:8_dig}, discussed in the previous section: (i) the DIG for not \hii\ hosts, (ii) the DIG for \hii\ hosts and (iii) the \hii\ regions. Note that this later component was already discussed in Sec. \ref{sec:results}.

\hii\ regions are clearly distributed at the location expected for those objects in the three diagrams. This is not a pure selection effect, as we already seen in Sec. \ref{sec:selHII} where we show that most \hii\ candidates were located in such regions before any further selection. 

More interesting are the differences in the distributions of the DIG for not \hii\ hosts and \hii\ hosts in the three diagrams. We should stress here that these two DIG components were selected based on the absence or presence of \hii\ regions in their host galaxies. Thus, no selection was done in any of the parameters involved in the three diagrams. In the BPT diagram the distribution for the DIG of the \hii\ hosts are located on top of the Kauffman curve, peaked at N2$\sim$-0.6 dex and O3$\sim$0.3 dex, covering a relatively narrow range of values in both parameters. On the WHAN diagram it is mostly located in the SF-region with a peak centered about log($|$EW(\Ha)$|$)$\sim$1.5 dex. Finally, in the WHAFY diagram it is distributed in a regime of intermediate EWs (above the usual limit of EW(\Ha)$>$6\AA\ adopted to select SF regions), but with lower values than the average one found for \hii\ regions in both parameters.

Finally, for the DIG corresponding to not \hii\ hosts a fraction is clearly located in the region corresponding to LINER-like/retired ionization regime and the other fraction is located in the AGN regime in the three diagrams (we suggest deep exploration of the nature of this double distribution to dismiss the effect of signal/noise cut-off applied): (i) well above the Kewley curve in the BPT diagram; (ii) in the retired region of the WHAN diagram, and (iii) well below Fyoung$<$4\% in the WHAFY diagram. The above is thus compatible with ionization due to HOLMES/post-AGB. With the present differences of the two types of DIG in the three diagrams, we reinforce the suggestion that the gas properties and the underlying (surrounding) stellar populations are different, and therefore produced by different physical processes.  

\section{Discussion}
\label{sec:discu}

In this paper we present the largest catalog of individual \hii\ regions ($\sim$52,000) obtained using IFU data based on a sub-sample of 678 galaxies extracted from the AMUSING++ compilation. Throughout this article we have shown that this sub-sample of galaxies, although heterogeneous, conforms a good representation of the galaxies in the nearby Universe (\citealp{califapresentation}). We also present the diffuse gas maps for these 678 galaxies derived as part of the analysis. For both the \hii\ region catalog and the diffuse ionized gas maps compendium we include the following information: (i) a set of parameters for 30 emission lines sampled by this wavelength range, including their fluxes, velocities, velocity dispersions and equivalent widths; and (ii) a set of properties of the underlying stellar populations, such as average ages and metallicities.

\subsection{Improvements in the derived observational properties}

The improvements in the exploration of the properties of \hii\ regions due to a high spatial resolution with respect to previous studies (i.e., MUSE data) have been already reported by various groups in the calculation of different properties, for instance: (i) to study the shape of oxygen abundance profiles \citep{sanchezmenguianosanchezperez2018}; (ii) to detect nebular lines and study the mechanisms ionizing in \hii\ regions \citep{mayyaplatgomez2023}; (iii) to find the effect in the SFR due to the presence of bars, ram pressure striping and/or nuclear activity \citep{sanchezcervantesfritz2023}; or (iv) to analyze the physics behind the dust content of nearby galaxies \citep{leesandstromleroy2023}. Unlike the direct precedent studies using large samples of IFS galaxies to explore the properties of \hii\ regions (\citealp{espi20, sanchezperezrosalesortega2015}), we are less affected by the spatial resolution as the data provided by MUSE may even be limited by natural seeing ($\lessapprox$1"arcsec). A more coarse resolution, as the one provided by the CALIFA data ($\approx$2.5"arcsec), affects the detectability of small regions that cannot be separated among themselves or from brighter adjacent ones. How this affect the observations and physical parameters of those \hii\ regions and galaxies in general was already discussed in previous studies (e.g., \citealp{mastrosalessanchez2014}). Furthermore, the use of \pyHII\ provides a decontamination of the contribution of the diffuse ionized gas for our catalog of \hii\ regions. This contamination has been studied in previous specific studies on star-forming galaxies (e.g., \citealp{mannuccibelfiorecurti2021}, \citealp{lacerdacidcouto2018}). 

Additional characteristics of the adopted tool are that: (i) it is able to recover \hii\ regions of a wide range of sizes, in contrast with other tools adopted in the analysis of low-spatial resolution IFS data where regions are recovered with similar sizes (e.g., {\sc HIIexplorer}, which was initially written in {\sc Perl} by \citealp{sanchezrosalesmarino2012} and transcribed to {\sc Python} by \citealp{espi20}. The diversity of sizes and the recovery of more regions in the low-luminosity regime would make this catalog better placed for the exploration of the \hii\ regions luminosity function \citep[e.g.,][]{santorokreckelbelfiore2022}; (ii) the modelling introduced by the code allows to decontaminate to a certain extent by the effects of the wings of spatial distribution of adjacent \hii\ regions and the PSF. This allows to separate between different adjacent regions and separate different ionizing sources such as the diffuse ionized gas component and the \hii\ regions \citep[which was not possible in previous explorations, e.g.,][]{espi20}. 

\subsection{Further ionizing sources that may pollute \hii\ regions}

Despite the above described improvements, the method adopted to select clumpy structures is not able to evade the contribution of the shocks due to outflows (\citealp{lopezcobasanchez2020}) and the ionization associated to the gas accretion of the supermassive black holes or AGNs (\citealp{lacerdasanchezcid2020}). Thus, both ionizations may impact somehow the results presented in this study (e.g., \citealp{daviesgroveskewley2017}). For instance a non negligible contribution of both components may affect the observed line ratios. In some of the panels included in Fig. \ref{fig:2}, there is an extension towards the LINER-line regime in both the BPT and WHAN diagram that could be associated with such pollution. 

Regarding the distribution in the WHAFY diagram, we notice that the fraction of young stars is clearly larger in the \hii\ region candidates than in the corresponding DIG. However, this fraction is still larger in the DIG of galaxies hosting \hii\ regions than in those that does not host them. This reflects the variety of physical mechanisms that generate the observed DIG in different galaxies and regions within them, comprising shocks ionization (\citealp{dopitasutherland1996}), post-AGB/HOLMES (\citealp{floresfajardomorissetstasinska2011}, \citealp{singhvandevenjahnke2013}) or photon leaked from \hii\ regions themselves (\citealp{relanokennicutteldridge2012}, \citealp{belfioresantorogroves2022}).

\subsection{Imprints of galaxy evolution on \hii\ regions}

Our exploration reinforces previous results suggesting a clear connection between the properties of \hii\ regions and the properties of their host galaxies, such as: (i) morphology; (ii) the integrated stellar mass; (iii) the galactocentric distance at which the \hii\ region is located; (iv) the LW age of the underlying stellar population and (v) to a lower degree the LW metallicity. From these results we conclude that this is reflected in the location in the diagnostic diagrams plus the intrinsic presence of ionizing OB-stars, that is dictated by the combination of all those galaxy properties. 

For instance, early-type/massive galaxies with more evolved and metal rich stellar populations (see Fig. \ref{fig:6_age} and Fig. \ref{fig:7_met}), have fewer \hii\ regions, and they are mostly located in the lower right-end of the BPT diagram. In late-type/less massive galaxies with more metal poor and younger stellar populations, the \hii\ regions are shifted towards the upper-left regime of the BPT diagram. This change in the location of the distributions is due to the different stellar populations (and properties) presents regarding their morphology and stellar mass of the galaxies, in accordance with the reported by previous studies, (e.g., \citealp{gonzalesgarciabenitoperez2015}). A similar pattern is observed when exploring the distributions regarding the galactocentric distance within the host galaxy (see Fig. \ref{fig:5_radii}). The center of galaxies is mainly populated by old, metal-rich stellar populations compared to the outer regions, whose populations are young and metal-poor. This gradient in the stellar populations (e.g., \citealp{goddardthomasmaraston2017, parikhthomasmaraston2021}) is indeed the responsible of the gradient observed in the oxygen abundance according with other studies (e. g. \citealp{belfioremaiolinotremonti2017}; 
\citealp{sanchezrosalesiglesias2014};
\citealp{sanchezsanchezmenguianomarino2015}; \citealp{sanchezmenguianosanchezperez2016}). This indeed supports the inside-out scenario for the formations of stellar component in galaxies (e.g., \citealp{pilkingtonfewgibson2012, zinchenkojustpilyugin2019}).  

These general trends have already been observed and discussed in \cite{espi20} and \cite{sanchezperezrosalesortega2015}. Nevertheless, our work, due to the large sample of \hii\ regions and the decontamination of the DIG, has allowed us point to make a more robust statistic to find more detailed trends. We found direct relations between the two explored line ratios (N2 and O3, see Fig. \ref{fig:10_age_zh}) and the average properties of the underlying stellar population (LW age and [Z/H]). In the case of N2, we find a moderate (weak) correlation with the age ([Z/H]). 

The physics behind the moderate and weak correlation of the N2 ratio and properties of the underlying stellar populations is due to the connection between the age of stellar populations, the production of metals throughout the life of the stars and enrichment in the middle of the galaxies. The most straight-forward explanation for this behavior is that the older stellar populations correspond to the most evolved ones, born in an epoch of the highest enrichment rate in galaxies (e.g., \citealp{valeasaricidstasinska2007}; \citealp{campssanchezlacerda2021}). Thus they corresponds to stellar populations with the highest metal content, that are located in areas in the which the ISM presents also the highest metal content (\citealp{tinsley1980}). Simultaneously, the N2 ratio is widely used as an tracer of the oxygen abundance. The connection between stellar metallicity of young stars and oxygen abundance was established in \cite{gonzalezcidgarcia2014}. So, if one increases the N2 ratio, one increases the age of the underlying stellar populations and the metallicity of them. In summary, based on the observed trends the oxygen abundance of the ISM is related to the star-formation and chemical enrichment history (\citealp{garciagonzalesperez2017, duartevilcheziglesias2022}). 

Based on our analysis of the line ratios and the properties of the underlying stellar populations summarized in Table \ref{tab:1}, we found that age dependence with the N2 line ratio is not strong in comparison with dependencies found in other studies using the CALIFA sample (e.g., \citealp{espi20}). This is maybe due to low reliability in the derivation of the properties of the underlying stellar populations due to limited wavelength range of MUSE in the blue range. However, in general correlation coefficients indicate a greater dependence of line ratios on age than on [Z/H]. 

On the other hand, for the O3 line ratio, there is an anti-correlation with the age and the metallicity of the underlying stellar populations. However, this ratio is more sensitive to other parameters such as the average ionization state, and the variation of the electron temperature, geometry, or density of the photoionized gas (\citealp{grovesheckmankauffmann2006, espi20}) and less sensitive to oxygen abundance. However, it is well known that this parameter anticorrelates with N2 in \hii\ regions, following the well known sequence along the BPT diagram described in Fig. \ref{fig:7_met}. Therefore, the anticorrelation between the stellar metallicity and O3 is most probably induced by the combined trends between O3 and N2, and between this later parameter and age. 

A particular case of study corresponds to the few \hii\ regions found in early-type galaxies (E/S0). These galaxies are generally absent of star-formation. Thus these \hii\ regions could be either remnants of a dimmed disk (e.g., in S0 galaxies), or the result of star formation induced by the capture of a gas-rich galaxies or pristine gas infall (\citealp{gomespapaderosvilchez2016}). The two scenarios may produce a different pattern in the properties of the ISM. In the first scenario, \hii\ regions should follow the same patterns as the ones described for late-type galaxies. However, in the second scenario, the properties of the \hii\ regions may be decoupled of that of the underlying stellar population. Our results suggest that a combination of both scenarios are required to explain the bimodality observed and trends in those bins corresponding to those morphological groups.

\subsection{The two natures of the DIG}

The DIG is a component that has already been studied on theoretical grounds using three-dimensional of radiative transfer simulations (e.g., \citealp{weberpauldrachhoffmann2019}) as well as from the observational point-of-view using large IFS dataset, such as MaNGA, CALIFA and MUSE  (e.g.,\citealp{valeasaricoutocidfernandes2019}, \citealp{lacerdacidcouto2018}, \citealp{denbrokcarolloerrozferrer2020}) or using CO molecular gas observations such as EDGE-CALIFA (e.g., \citealp{levybolattosanchez2019}). However, as mentioned above, the considerable improvement in the spatial resolution provided by our data opens new perspective in the exploration of this emission component.

Our explorations suggest in Sec. \ref{sec:dig_types} that the distribution of EW(\Ha) for the diffuse component in our galaxies is more similar to the one reported for the CALIFA sample \citep{lacerdacidcouto2018}, than the one more recently reported based on the PHANGS-MUSE sample \citep{belfioresantorogroves2022}. Our physical spatial resolution is clearly not as good as the one provided by the PHANGS-MUSE sample, but it is considerable better than the one provided by CALIFA. Thus, spatial resolution is not the main driver for this results. We consider that the fact that PHANGS-MUSE sample is essentially biased towards late-type galaxies is indeed the main driver for the reported difference, as the distributions for the DIG found for this sub-sample of galaxies in the AMUSING++ sample does indeed agree with the one reported by PHANGS-MUSE.

The distribution of EW(\Ha) for the spatial resolved DIG in our complete sample of galaxies presents a clear bimodal shape. It presents two well defined peaks centered at $\sim$1\AA\ and $\sim$10\AA, which according to the classification by \cite{lacerdacidcouto2018}, corresponds to hDIG ionization (component dominated by photoionization by old, low-mass stars and evolved or called HOLMES), and the ionization of SF complexes (component dominated by unresolved \hii\ region, or photons leaked from them). These two kind of diffuse ionization have been widely studied, in previous explorations (e.g., \citealp{zuritabeckmanrozas2002}, \citealp{floresfajardomorissetstasinska2011} and \citealp {weilbachermonrealiberoverhamme2018}).  

To explore in detail the nature of the two main diffuse components found when separating galaxies between \hii\ region hosts and not hosts, we show in Figure \ref{fig:9_diagrams} their distribution in the three diagnostic diagrams adopted along this article: BPT, WHAN and WHAFY. The location of the \hii\ region candidates has been included in that figure too. By construction the later component is located in the expected regimes: (i) below the Kewley line in the BPT diagram; (ii) at the SF area in the WHAN diagram, and (iii) in the regime of high EW(\Ha) and fraction of young stars in the WHAFY diagram. 

On the other hand, both diffuse components cover a different regime in the diagrams, in particular the diffuse found in not \hii\-regions hosting galaxies, a component that is located either in the LINER-like area in the three diagrams, with a fraction of these diffuse found in the AGN regime. Based on these results we propose that this DIG is mostly due to ionization processes caused by old stellar populations, with a possible contribution of AGN or shocks. This would correspond to the ionization reported in the extraplanar areas of edge-on spirals (\citealp{floresfajardomorissetstasinska2011}, \citealp{lulivargas2023}), in the bulge and inter-arm regions of face-on spirals \citep{singhvandevenjahnke2013}, and in early-type galaxies in general \citep{gomespapaderosvilchez2016}. The distribution of the DIG of galaxies hosting \hii\ regions is clearly different from this one in the three diagrams. Indeed, it presents values more similar to those reported for \hii\ regions themselves, but with evidence of harder ionization (i.e., larger values of both O3 and N2).
Based on these results we consider that this ionization is most probably the result of the combination of unresolved low-luminous \hii\ regions and the presence of photons leaked from the adjacent \hii\ regions (e.g. {\it leaking}, \citealp{fergusonwysegallagher1996}). This double nature of the diffuse has been recently explored by \citet{belfioresantorogroves2022} with similar conclusions, using PHANGS-MUSE data.  

Another contribution that may be an important source of ionization in the DIG is the one due to shocks of high- and low-velocity \citep[e.g.][]{dopitasutherland1996}. Indeed, this component may be present in galaxies of any morphological type. It is important to note that the presence of shocks may enhance low-ionization forbidden lines, as observed in the DIG of certain galaxies (\citealp{hofilippenkosargent1997}). Explorations including other parameters, such as the velocity dispersion (\citealp{dagostinokewleygroves2019}; Sanchez et al. in prep) and/or detailed ionization modelling (e.g., \citealp{sutherlanddopita2017} and \citealp{alariemorisset2019}) may be required to determine its real contribution to the ionization budget in DIG.

\section{Conclusions}
\label{sec:conclu}

In this work we have presented the largest catalog of spectroscopic properties of \hii\ region extracted from a large sample of galaxies covering a fairly representative range of parameters (mass, colors, morphologies) of the population in the nearby Universe. We explored the main distributions of those \hii\ regions in a set of diagnostic diagrams, segregating them by the properties of their (i) host galaxies; (ii) the underlying stellar populations, and (iii) the location within galaxies. From our exploration we want to highlight the following:

\begin{itemize}
\item The \hii\ regions in E\&S0 galaxies do not follow the trends described for the remaining morphological types for all the explorer properties. 
\item For the rest of the morphological types, \hii\ regions in (i) more massive galaxies, (ii) central areas (i.e., the bulge), with (iii) old ages and/or (iv) high metallicities of the underlying stellar populations are located in the lower-right areas of the BPT diagrams with low values of EW(\Ha). 
\item On the other hand, (i) as the stellar mass of the host galaxy is lower, (ii) at greater galactocentric distances (disk area), with (iii) younger ages and/or (iv) lower metallicities of the underlying stellar population, \Hii\
are located towards the upper-left areas of the BPT diagrams with higher values of EW(\Ha).
\end{itemize}

We report on the deep connection between the observational and physical properties of those ionized regions and the explored parameters, mainly the age of the underlying stellar populations with the line ratios N2 (correlation) and O3 (anti-correlation). Finally, we describe the EW(\Ha) values for (i) the diffuse ionized gas of four galaxies along the Hubble sequence, (ii) our two main ionized components from which the catalogue of \hii\ regions is derived (candidates and DIG) and (iv) the double-nature of the diffuse ionize gas, clearly segregated in galaxies hosting and not hosting \hii\ regions.

Previous explorations found similar results and trends than those discussed here. However, we should stress out that the current analysis is performed on one of the most statistically robust and large sample of \hii\ regions, that make use of a state-of-the art methodology and high-spatial resolution and quality IFS data. This approach has allowed us to perform one of the most sophisticated decontamination by the elusive diffuse ionized gas contribution.
 
The analysis presented here opened new paths for further studies: (i) the exploration of the physical properties included in the catalogue of \hii\ regions (for instance, those related to size and absolute luminosity, or using line ratios for calculate electron temperature, density or abundance of oxygen in gas phase); (ii) the seek for ``anomalous" \hii\ regions (i.e., regions with peculiar properties, either observational or physical ones); and (iii) the exploration of the spatial resolved properties of the diffuse ionized gas. 
We will address them upcoming in studies.

\section*{Acknowledgements}


A.Z.L.A. and S.F.S. thank the PAPIIT-DGAPA AG100622 project. J.K.B.B. and S.F.S. acknowledge support from the CONACYT grant CF19-39578. 

L.G. acknowledges financial support from the Spanish Ministerio de Ciencia e Innovaci\'on (MCIN), the Agencia Estatal de Investigaci\'on (AEI) 10.13039/501100011033, and the European Social Fund (ESF) ``Investing in your future" under the 2019 Ram\'on y Cajal program RYC2019-027683-I and the PID2020-115253GA-I00 HOSTFLOWS project, from Centro Superior de Investigaciones Cient\'ificas (CSIC) under the PIE project 20215AT016, and the program Unidad de Excelencia Mar\'ia de Maeztu CEX2020-001058-M.

J.P.A. acknowledges financial support from ANID, Millennium Science Initiative, ICN12\_009.

This research made use of
Astropy,\footnote{http://www.astropy.org} a community-developed core {\sc Python} package for Astronomy \citep{astropy:2013, astropy:2018}. Public data obtained from the ESO Science Archive Facility with different DOIs. 

\section*{Data Availability}

The original MUSE datacubes are available in ESO archive (\url{https://archive.eso.org/wdb/wdb/adp/phase3_main/form}), however the processed MUSE datacubes by the \pipe\ pipeline or dataproducts used to obtain the catalogue of \hii\ regions will be shared on reasonable request to the corresponding author \citep{lopezcobasanchez2020}. The complete catalogue of \hii\ regions is public and is available in: \url{http://ifs.astroscu.unam.mx/MUSE/pyHIIexplorer/list_new.php}. On the other hand, the DIG maps of the all galaxies (678) and the complete catalogue of candidate \hii\ regions will be shared on reasonable request to the corresponding author for this work.

\bibliographystyle{mnras}
\bibliography{my_bib_HII_regions_catalogue} 

\appendix

\section{Description of catalogue}
\label{sec:desc_cat}

The catalogue of \hii\ regions described in this work is public and free access: \url{http://ifs.astroscu.unam.mx/MUSE/pyHIIexplorer/}. We present all derived parameters for both the emission lines and the underlying stellar populations of each individual \hii\ region (for each galaxy) in four different files in the ECSV format \footnote{\url{https://docs.astropy.org/en/stable/api/astropy.io.ascii.Ecsv.html}}: (i) .flux\_elines.ecsv; (ii) .SSP.ecsv; (iii) .SFH.ecsv and (iv) .INDEX.ecsv. To identify each region, we include in each file the following parameters: the \hii\ region ID, position and radius in pixels together with the right ascension and declination in degrees.

Below we briefly describe the contents of each file. 

(i) \texttt{HIIREGIONSID.galaxy\_name.flux\_elines.ecsv}: File that includes the fluxes, velocities, equivalent widths, velocities dispersions and errors of each derived parameter for each \hii\ region.

(ii) \texttt{HIIREGIONSID.galaxy\_name.SSP.ecsv}: File that contains the average properties of the underlying stellar populations derived by \pipe\ \citep{sanchezperezsanchez2016fit3d, lacerdasanchezmejia2022} including the luminosity- and mass-weighted age and metallicity, the dust attenuation, stellar mass among others (see \citealp{lacerdasanchezmejia2022}). 

(iii)
\texttt{HIIREGIONSID.galaxy\_name.SFH.ecsv}:
File that includes the fraction of light in the V-band corresponding to each template in the adopted SSP library.

(iv) \texttt{HIIREGIONSID.galaxy\_name.INDEX.ecsv}: File that contains the average properties of a set of stellar absorption indices derived once subtracted the emission line contribution. Such as the: H$_{\textit{$\delta$}}$, H$_{\textit{$\gamma$}}$, Fe5270 and Fe5335n (Lick/IDS index system, \citealp{bursteinfabergaskell1984}).   

Additionally we added four images for visual interpretation of the user: (i) gri three color continuum image; (ii) [OIII], \Ha\, [NII] three color image; (iii) \pyHII\ summary and (iv) BPT and WHAN diagrams. The description of each image is include in the webpage: \url{http://ifs.astroscu.unam.mx/MUSE/pyHIIexplorer/}.
  
\section{Effects of the decontamination by the contribution of the DIG}
\label{sec:dec_dig}

Throughout this study, we present a catalogue of \hii\ regions whose properties are decontaminated by the contribution of the DIG. In this appendix, we show the importance of the adopted decontamination in the emission line properties. We recall that the decontamination is performed by \pyHII\ itself (for more details on the procedure see sections 3.2, 3.3 and 3.4 of the paper \citet{lugosanchezespinosa2022}), using the same procedure for all galaxies as part of the detection and extraction of the properties of the clumpy regions. On the contrary, recent explorations did not apply any decontamination at all \citep{kreckelhoblanc2019}, the applied one is purely statistical \citep{espi20} or based on assumed properties of the DIG as \sii/\Ha\ and \Ha\ surface brightness \citep{zhangyanbyndy2017}.  

For comparison purposes, we derived an additional catalog of \hii\ regions without decontaminating due to the DIG. For doing so, we modified the \pyHII\ code removing the DIG decontamination. We repeated all the procedure of detecting and classifying \hii\ regions described along this article using this version of the code.

\begin{figure}
    \minipage{0.5\textwidth}
    \includegraphics[width=\linewidth]
    {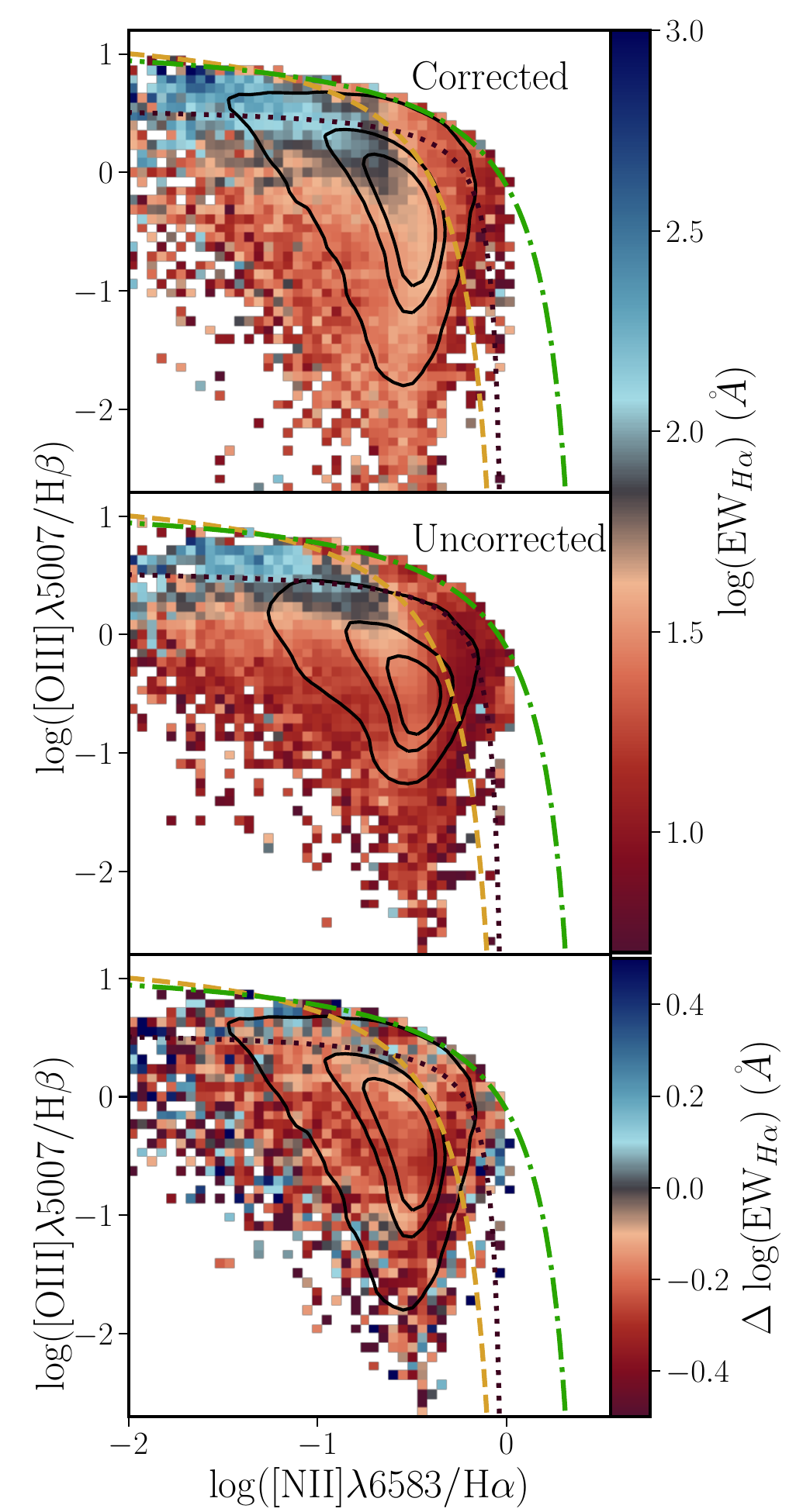}
    \caption{BPT diagram using the same nomenclature as the one adopted in Fig. \ref{fig:4_hii}-\ref{fig:7_met}, for the final sample of \hii\ regions corrected (52371 \hii\ regions {\it top panel}) and uncorrected (48649 \hii\ regions {\it middle panel}) for the contribution of the DIG. The difference of both distributions is shown in the {\it bottom panel}, with the contours being the same as the one shown in the top panel.}
    \endminipage
    \label{fig:12_dig}
\end{figure}

Figure \ref{fig:12_dig} shows the distribution of both catalogues along the classical BPT diagram, decontaminated (top panel) and non-decontaminated (middle panel) by the DIG contribution, as well as the difference between the average EWs(\Ha) (bottom panel). 

The main differences to be highlighted between both catalogs are: (i) in average the decontaminated \hii\ regions have log($|$EW(\Ha)$|$)$=$1.5 dex values while that in average the contaminated regions have log($|$EW(\Ha)$|$)$=$1.3 dex values; (ii) the distribution of the decontaminated catalog extends in the BPT diagram towards higher O3 and lower N2 values; (iii) in the catalog contaminated by DIG, we recover a smaller number of \hii\ regions (48649) compared to the regions that we recover uncontaminated due to DIG contribution (52371). The optimal recovery is not precisely that of the largest number of regions, this particular point we propose could be due to the DIG contribution that can decrease the signal to noise ratio; (iv) finally, the contours of density follow better the classical location of \hii\ regions in the case of the decontaminated catalog. Regarding the difference in the average EWs shown in the bottom panel, it is seen that the contribution of the DIG is homogeneous across the BPT diagram, irrespective of the main properties determining the location in this diagram (such as the oxygen abundance). 

These results suggest that the contamination by the DIG should not be ignored when exploring the properties of \hii\ regions at the current spatial resolution. 

\begin{figure*}
    \minipage{0.85\textwidth}
    \includegraphics[width=\linewidth]
    {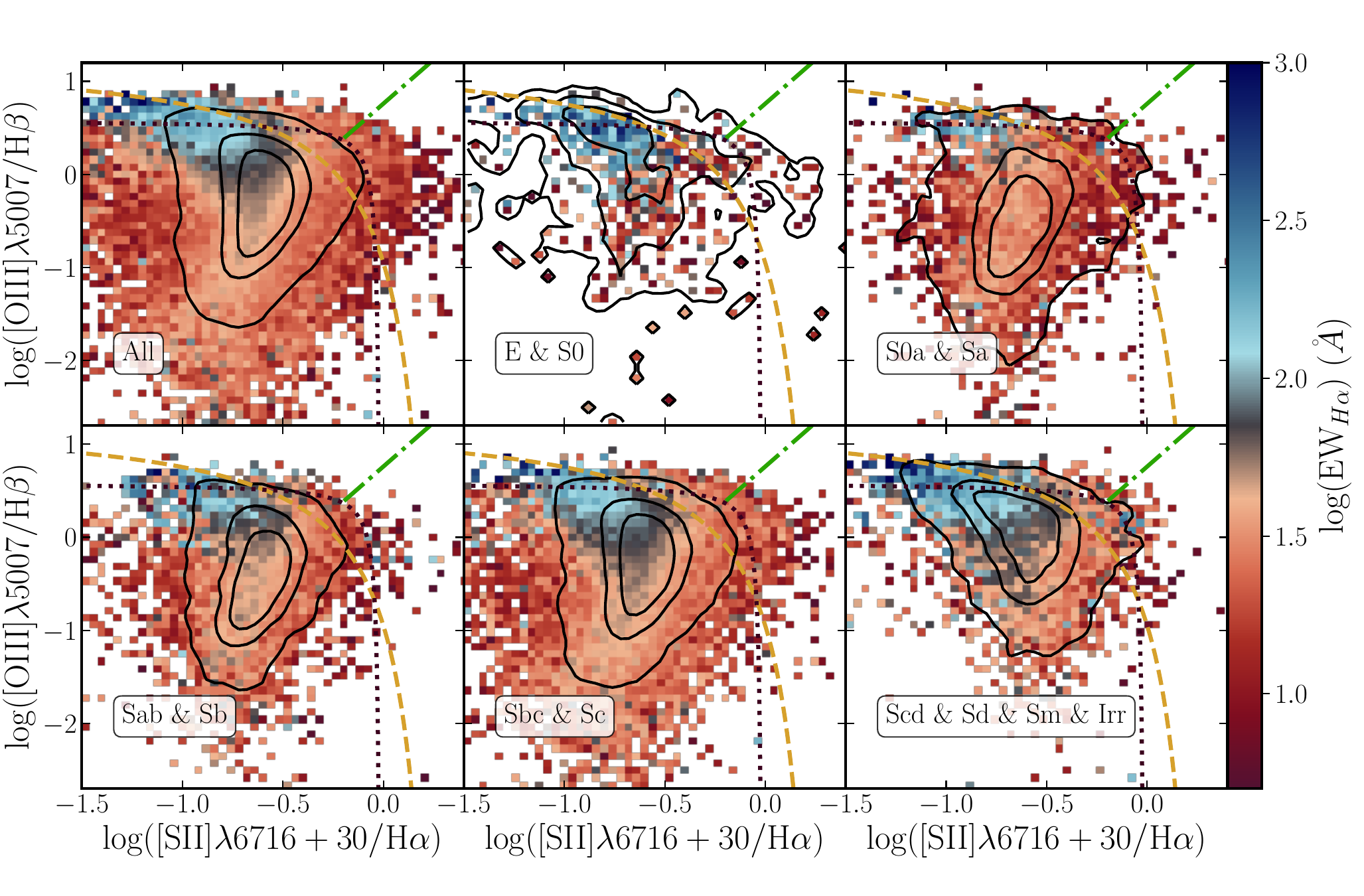}
    \endminipage
    \caption{\oiii$\lambda$5007/\Hb\ versus \sii$\lambda$6716+30/\Ha\ diagnostic diagram for our final sample of \hii\ regions segregated by morphology of their respective host galaxies. From left to bottom and top to bottom: the galaxies are grouped in panels like the columns shown in Fig. \ref{fig:4_hii}. Black-dotted, yellow dashed and green dot-dashed lines represent \citet{espi20} and \citet{kewleygroveskauffmann2006} demarcation lines. Colormaps and density contours are same as Fig. \ref{fig:4_hii}. }
    \label{fig:S2_O3}
\end{figure*}

\begin{figure*}
    \minipage{0.85\textwidth}
    \includegraphics[width=\linewidth]
    {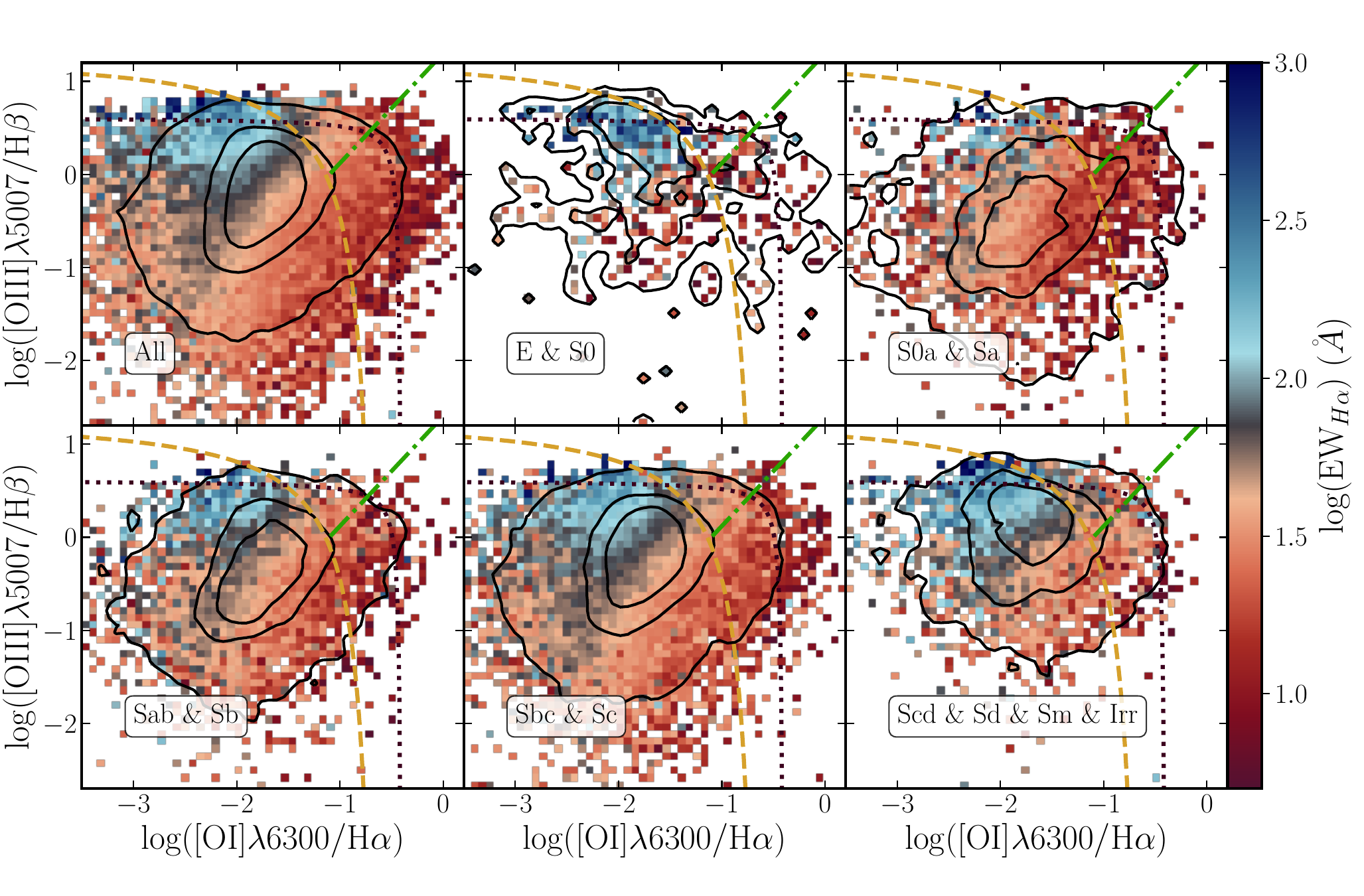}
    \endminipage
    \caption{\oiii$\lambda$5007/\Hb\ versus \oi$\lambda$6300/\Ha\ diagnostic diagram for our final sample of \hii\ regions segregated by morphology of their respective host galaxies. From left to bottom and top to bottom: the galaxies are grouped in panels like the columns shown in Fig. \ref{fig:4_hii}. Black-dotted, yellow dashed and green dot-dashed lines represent \citet{espi20} and \citet{kewleygroveskauffmann2006} demarcation lines. Colormaps and density contours are same as Fig. \ref{fig:4_hii}.}
    \label{fig:O1_O3}
\end{figure*}

\section{Other line ratios (S2 and O1)}

Along this paper we use the classical BPT diagram involving O3 and N2 \citep{baldwin81} to explore the properties of \hii\ regions in our catalog. In this appendix we explore the distributions along other frequently used diagnostics diagrams: [OIII]$\lambda$5007/\Hb\ (O3) versus [SII]$\lambda$6716+30/\Ha\ (S2), in Fig. \ref{fig:S2_O3} and [OIII]$\lambda$5007/\Hb\ (O3) versus [OI]$\lambda$6300/\Ha\ (O1), in Fig. \ref{fig:O1_O3} \citep{veilleuxosterbrockdonald1987}. In both figures we show the density distributions and the average EWs for all galaxies segregated by morphology.
 
In general, for figures \ref{fig:S2_O3} and \ref{fig:O1_O3} we observe that the \hii\ regions in our catalog fall below the demarcation lines usually adopted to select \hii\ regions in so-called star-forming or \hii\ regions zones. In more detail, the contours of the diagram O3 versus S2 fit more into the location of classical \hii\ regions in comparison with the distribution seen in the diagram O3 versus O1. We observe trends regarding the morphological type in both figures, the \hii\ regions present in the intermediate galaxies such as Sa and Sb are located in the lower right, while that in late-type galaxies such as Sc, Sd and Irr are located in the upper left with high values of EW(\Ha). Finally, the distributions for early-type galaxies (E \& S0), do not follow the trend appreciated from late-to-intermediate type spirals. On the contrary, \hii\ regions in these galaxies are located at the top-right end of the distribution, consistently with what it was found for the classical BPT diagram. This agrees with what we suggest and comment in previous sections (Sec. \ref{sec:results} and \ref{sec:discu}) is due to the rejuvenation due to the inflow of more metal poor gas or the capture of low-mass/low-metallicity galaxies. 

\end{document}